\documentstyle[multicol,aps,prl,psfig,array]{revtex}
\renewcommand{\narrowtext}{\begin{multicols}{2} \global\columnwidth20.5pc}
\renewcommand{\widetext}{\end{multicols} \global\columnwidth42.5pc}
\def\al{\alpha}
\def\be{\beta}
\def\ga{\gamma}
\def\de{\delta}
\def\ep{\epsilon}
\def\ve{\varepsilon}
\def\ze{\zeta}
\def\et{\eta}
\def\th{\theta}

\def\ka{\kappa}
\def\la{\lambda}

\def\rh{\rho}

\def\si{\sigma}
\def\vs{\varsigma}

\def\ph{\phi}

\def\ch{\chi}
\def\ps{\psi}
\def\om{\omega}

\def\De{\Delta}

\def\La{\Lambda}
\def\Si{\Sigma}

\def\Ph{\Phi}

\def\Om{\Omega}
\def\mn{{\mu\nu}}
\def\cl{{\cal L}}
\def\fr#1#2{{{#1} \over {#2}}}
\def\prt{\partial}

\def\pt#1{\phantom{#1}}

\def\expect#1{\langle{#1}\rangle}

\def\half{{\textstyle{1\over 2}}}
\def\frac#1#2{{\textstyle{{#1}\over {#2}}}}

\def\lsim{\mathrel{\rlap{\lower4pt\hbox{\hskip1pt$\sim$}}
    \raise1pt\hbox{$<$}}}
\def\gsim{\mathrel{\rlap{\lower4pt\hbox{\hskip1pt$\sim$}}
    \raise1pt\hbox{$>$}}}
\def\sqr#1#2{{\vcenter{\vbox{\hrule height.#2pt
         \hbox{\vrule width.#2pt height#1pt \kern#1pt
         \vrule width.#2pt}
         \hrule height.#2pt}}}}

\def\Re{\hbox{Re}\,}
\def\Im{\hbox{Im}\,}

\def\curl#1{\vec\nabla\times\vec #1}
\def\div#1{\vec\nabla\cdot\vec #1}
\def\grad#1{\vec\nabla #1}

\newcommand{\beq}{\begin{equation}}
\newcommand{\eeq}{\end{equation}}
\newcommand{\bea}{\begin{eqnarray}}
\newcommand{\eea}{\end{eqnarray}}
\newcommand{\rf}[1]{(\ref{#1})}

\def\kaf{k_{AF}}
\def\kf{k_{F}}
\def\kfi{(k_{F})_{\ka\la\mu\nu}}
\def\kt{\tilde k}
\def\pht{\hat p}
\def\sind{\sin d} \def\cosd{\cos d}
\def\sinr{\sin r} \def\cosr{\cos r}
\def\sinds{\sin^2 d} \def\cosds{\cos^2 d}
\def\sinrs{\sin^2 r} \def\cosrs{\cos^2 r}

\begin{document}

\title{Signals for Lorentz Violation in Electrodynamics}
\author{V.\ Alan Kosteleck\'y and Matthew Mewes}
\address{Physics Department, Indiana University,
Bloomington, IN 47405, U.S.A.}
\date{IUHET 449, May 2002}

\maketitle

\begin{abstract}
An investigation is performed of
the Lorentz-violating electrodynamics 
extracted from the renormalizable sector of the
general Lorentz- and CPT-violating standard-model extension.
Among the unconventional properties of radiation
arising from Lorentz violation is birefringence of the vacuum.
Limits on the dispersion of light  
produced by galactic and extragalactic objects
provide bounds of $3\times 10^{-16}$
on certain coefficients for Lorentz violation in the photon sector.
The comparative spectral polarimetry of light 
from cosmologically distant sources
yields stringent constraints of $2\times 10^{-32}$.
All remaining coefficients in the photon sector
are measurable in high-sensitivity tests
involving cavity-stabilized oscillators.
Experimental configurations in Earth- and space-based laboratories
are considered that involve optical or microwave cavities 
and that could be implemented using existing technology.
\end{abstract}


\narrowtext

\section{INTRODUCTION}

Lorentz symmetry underlies the theory of relativity 
and all accepted theoretical descriptions of nature
at the fundamental level.
A crucial role in establishing 
both the rotation and boost components of 
Lorentz symmetry has been played by
experimental studies of the properties of light.
In the classic tests, 
rotation invariance is investigated 
in Michelson-Morley experiments
searching for anisotropy in the speed of light,
while boost invariance is studied 
via Kennedy-Thorndike experiments
seeking a variation of the speed of light
with the laboratory velocity
\cite{mm,kt,classic}.

In this work,
a theoretical study is performed of various experiments testing
Lorentz symmetry with light and other electromagnetic radiation.
The analysis is within the context of 
the Lorentz- and CPT-violating standard-model extension
\cite{ck},
developed to allow for small general violations 
in Lorentz and CPT invariance
\cite{cpt01}.
The lagrangian of this theory includes
all observer Lorentz scalars formed by combining standard-model fields
with coupling coefficients having Lorentz indices.
At the level of quantum field theory,
the violations can be regarded as remnants of Planck-scale physics 
appearing at attainable energy scales.
The coefficients for Lorentz violation may be
related to expectation values of Lorentz tensors or vectors
in an underlying theory
\cite{kps}.
To date, 
experimental tests of the standard-model extension
have been performed with hadrons
\cite{kexpt,bexpt,dexpt,kpcvk},
protons and neutrons
\cite{ccexpt},
electrons 
\cite{eexpt,eexpt2},
photons
\cite{cfj,km},
and muons
\cite{muexpt}.

In the present context of studies of electrodynamics,
the standard-model extension is of interest
because it provides a general field-theoretic framework for 
investigating the Lorentz properties of light.
The theory contains as a subset
a general Lorentz-violating 
quantum electrodynamics (QED),
which includes a general Lorentz-violating extension
of the Maxwell equations.
We study experiments that can measure 
the coefficients for Lorentz violation 
in this generalized electrodynamics.
Our attention is restricted here to 
exceptionally sensitive experiments
that could be in a position to detect the
minuscule effects motivating the standard-model extension.

A basic feature of Lorentz-violating electrodynamics
is the birefringence of light propagating 
\it in vacuo. \rm
This results in several potentially observable effects,
including pulse dispersion and polarization changes.
One goal of this work
is to consider the implications of these effects
for the propagation of radiation on astrophysical scales.
We use available observations to constrain
certain coefficients for Lorentz violation. 

Another goal of this work is to analyze 
modern versions of some classic tests 
of special relativity
based on resonant-cavity oscillators
\cite{bh,hh,brax},
which have extreme sensitivity to the properties
of electromagnetic fields.
These experiments depend on the Earth's sidereal and orbital motion.
However,
the advent of the International Space Station (ISS)
makes it feasible to perform laboratory experiments in space,
where the orbital motion can yield different sensitivity
to Lorentz-violating effects
\cite{bklr}.
We consider here both space- and Earth-based laboratory experiments
with resonant cavities.

The structural outline of this paper is as follows.
Section \ref{lve}
presents some basic results and definitions for 
the general Lorentz-violating electrodynamics
and outlines the connection to some test models.
We then consider birefringence experiments,
beginning in Sec.\ \ref{biretheory} 
with some general issues.
Constraints stemming from the resulting effects
on pulse dispersion from astrophysical sources
are addressed in Sec.\ \ref{velocity},
while those from polarization changes over
cosmological scales are treated in Sec.\ \ref{polar}. 
A general analysis for laboratory-based experiments
on the Earth and in space is presented in
Sec.\ \ref{labtheory}.
Sections \ref{optcav} and \ref{miccav} apply this
analysis to experiments
with optical and microwave resonant cavities.
We summarize in Sec.\ \ref{summary}.
Throughout this work,
we adopt the conventions of Ref.\ \cite{ck}.

\section{LORENTZ-VIOLATING ELECTRODYNAMICS}
\label{lve}

This section provides some background 
and contextual information
about the general Lorentz-violating electrodynamics.
The basic formalism is presented,
and some definitions used in later sections are introduced.
We also discuss the connection between this theory
and some test models for Lorentz violation.

\subsection{Basic Theory}
\label{theory}

The standard model of particle physics is believed to
be the low-energy limit of a fundamental theory that
includes all the forces in nature.
The natural scale of this fundamental theory is likely to be
determined by the Planck mass.
The possibility that 
Lorentz- and CPT-violating signals from this theory
may be observable at energies attainable today 
led to the development of the standard-model extension
\cite{ck},
which is a general theory based on the standard model
but allowing for violations of Lorentz and CPT symmetry
\cite{cpt01}.
The additional terms must be small
because the usual standard model agrees well with experiment.
They may originate from
spontaneous symmetry breaking in the fundamental theory
\cite{kps}.

The standard-model extension can be defined as
the usual standard-model lagrangian plus
all possible additional Lorentz- and CPT-violating terms 
involving standard-model fields
that maintain invariance under
Lorentz transformations of the observer's inertial frame.
This invariance ensures that the physics
is independent of the choice of coordinates.
The Lorentz violation is associated with 
rotations and boosts of particles or
localized field configurations in a fixed observer inertial frame. 

Many of the detailed investigations of the standard-model extension
have been performed under the simplifying assumption
that the additional Lorentz- and CPT-violating terms
preserve the SU(3)$\times$SU(2)$\times$U(1)
local gauge symmetry of the usual standard model.
Another widely adopted simplifying assumption
is that the coefficients for Lorentz violation
are independent of position.
This implies the violation is restricted to the 
Lorentz symmetry instead of the full Poincar\'e symmetry
and has several useful consequences for experiment,
including the conservation of energy and momentum.
It is also often convenient to restrict attention
to the renormalizable sector of the theory,
since this is expected to dominate the physics at low energies.
However,
nonrenormalizable terms are known to play 
an important role at higher energies
\cite{kle}.

Extracting terms involving the photon fields from the
standard-model extension yields 
a Lorentz- and CPT-violating extension of QED 
\cite{ck}.
The fermion sector of this theory has been widely studied.
Here,
we focus attention on the pure-photon sector
and limit attention to the renormalizable terms,
which involve operators of mass dimension four or less.
The relevant lagrangian is \cite{ck}
\bea
\cl & = & -\frac 1 4 F_{\mu\nu}F^{\mu\nu}
+\frac 1 2 (\kaf)^\ka\ep_{\ka\la\mu\nu}A^\la F^{\mu\nu}
\nonumber \\
& & 
- \frac 1 4 \kfi F^{\ka\la}F^{\mu\nu} ,
\label{lagrangian}
\eea
where $F_\mn \equiv \prt_\mu A_\nu -\prt_\nu A_\mu$.
This theory maintains the usual U(1) gauge invariance
under the transformations $qA_\mu \to qA_\mu + \prt_\mu \La$.
The lagrangian contains the standard Maxwell term 
and two additional Lorentz-violating terms.
The first of these extra terms is CPT odd,
and its coefficient $(\kaf)^\ka$ has dimensions of mass.
The other is CPT even.
Its coefficient $\kfi$ is dimensionless
and has the symmetries of the Riemann tensor 
and a vanishing double trace,
which implies a total of 19 independent components.

The CPT-odd term has received much attention
in the literature \cite{jk}.
This term provides negative contributions to the canonical energy 
and therefore is a potential source of instability.
One solution is to set the coefficient to zero,
$(\kaf)^\ka=0$.
This is theoretically consistent with radiative corrections
in the standard-model extension
and is well supported experimentally:
stringent constraints on $\kaf$ 
have been set by studying the polarization of radiation
from distant radio galaxies
\cite{cfj}.

In contrast,
much less is known about the CPT-even coefficient $\kf$.
Theoretical studies show that it provides
positive contributions to the canonical energy
and that it is radiatively induced from the fermion sector 
in the standard-model extension
\cite{ck,klp}.
Constraints on some components have recently been
obtained from optical spectropolarimetry of
cosmologically distant sources
\cite{km}.
In the present work,
we focus on the experimental implications of this CPT-even term.
The coefficient $(\kaf)^\ka$ is set to zero for the analysis.

The equations of motion from lagrangian \rf{lagrangian} are 
\beq
\prt_\al{F_\mu}^\al+(\kf)_{\mu\al\be\ga}\prt^\al F^{\be\ga}=0 .
\label{max1}
\eeq
These are modified source-free inhomogeneous Maxwell equations.
The homogeneous Maxwell equations,
\beq
\prt_\mu \widetilde F^{\mn}
\equiv \half\ep^{\mu\nu\ka\la}\prt_\mu F_{\ka\la}=0,
\label{max1hom}
\eeq
remain unchanged.

Although it lies beyond our present scope,
the techniques presented here and the results obtained
can be generalized to the nonrenormalizable sector.
The nonrenormalizable terms can be classified
according to their mass dimension.
The dimensions of the corresponding coefficients 
are inverse powers of mass,
and it is plausible that  
these coefficients are suppressed by corresponding powers
of the Planck scale.
Terms of this type appear in various special Lorentz-violating theories,
including noncommutative field theories 
incorporating QED 
\cite{chklo}.
Indeed, 
any coordinate-independent theory with a photon sector containing  
nonrenormalizable Lorentz-violating terms
must be a subset of the standard-model extension.
It would be interesting to  
provide a detailed study of the nonrenormalizable terms in the
Lorentz-violating electrodynamics
and their experimental signals.

\subsection{Analogy and Definitions}
\label{defs}

A useful analogy exists between the Lorentz-violating electrodynamics 
\it in vacuo \rm 
and the conventional situation 
in homogeneous anisotropic media
\cite{ck}.
The idea is to \it define \rm fields 
$\vec D$ and $\vec H$
by the six-dimensional matrix equation
\beq
\left(
\begin{array}{c} 
\vec D \\ \vec H 
\end{array} 
\right)
=
\left(
\begin{array}{cc} 
 1+\ka_{DE} & \ka_{DB} \\
 \ka_{HE} & 1+\ka_{HB} 
\end{array}
\right)
\left(
\begin{array}{c} 
\vec E \\ \vec B 
\end{array} 
\right) ,
\label{DH}
\eeq
where $\vec E$ and $\vec B$ are the electric and magnetic fields 
obtained from solving the modified Maxwell equations \rf{max1}.
The $3 \times 3$ matrices
$\ka_{DE}$, $\ka_{HB}$, $\ka_{DB}$,
and $\ka_{HE}$ 
are defined by
\bea
(\ka_{DE})^{jk} &=& -2 (\kf)^{0j0k}, 
\nonumber \\
(\ka_{HB})^{jk} &=& \half \ep^{jpq} \ep^{krs} (\kf)^{pqrs}, 
\nonumber \\
(\ka_{DB})^{jk} &=& -(\ka_{HE})^{kj} = (\kf)^{0jpq}\ep^{kpq}.
\label{kappas}
\eea
The double-trace condition on $\kfi$ 
translates to the tracelessness of
$(\ka_{DE}+\ka_{HB})$,
while $(\kf)_{\ka[\la\mu\nu]}=0$ implies 
the tracelessness of $\ka_{DB} = -(\ka_{HE})^T$.
This leaves $\ka_{DE}$ and $\ka_{HB}$
with eleven independent elements
and the matrix $\ka_{DB} = -(\ka_{HE})^T$ 
with eight,
which together represent 
the 19 independent components of $\kf$.
Note also that $\ka_{DE}$ and $\ka_{HB}$ are
parity even, 
while $\ka_{DB} = -(\ka_{HE})^T$ is parity odd.

With these definitions, 
the modified Maxwell equations \rf{max1}, \rf{max1hom}
take the familiar form
\bea
\curl H - \prt_0 {\vec D} = 0, & &\quad \div D = 0, \nonumber \\
\curl E + \prt_0 {\vec B} = 0, & &\quad \div B = 0.
\label{max2}
\eea
As a consequence, 
many results from conventional electrodynamics 
in anisotropic media also hold for this Lorentz-violating theory.
For example,
the energy-momentum tensor takes the standard form in terms
of $\vec E$, $\vec B$, $\vec D$ and $\vec H$.
This implies the usual Poynting theorem,
which can be applied in conjunction
with the symmetries of the matrices in
Eq.\ \rf{DH} to show that the vacuum is lossless.

For the applications to be addressed in later sections,
it is convenient to introduce the following
decomposition of $\kfi$ coefficients:
\bea
(\tilde\ka_{e+})^{jk}&=&\half(\ka_{DE}+\ka_{HB})^{jk},
\nonumber\\
(\tilde\ka_{e-})^{jk}&=&\half(\ka_{DE}-\ka_{HB})^{jk}
                 -\frac13\de^{jk}(\ka_{DE})^{ll},
\nonumber\\
(\tilde\ka_{o+})^{jk}&=&\half(\ka_{DB}+\ka_{HE})^{jk} ,
\nonumber\\
(\tilde\ka_{o-})^{jk}&=&\half(\ka_{DB}-\ka_{HE})^{jk},
\nonumber\\
\tilde\ka_{\rm tr}&=&\frac13(\ka_{DE})^{ll}.
\label{kappas2}
\eea
The first four of these equations define 
traceless $3\times3$ matrices,
while the last defines a single coefficient.
All parity-even coefficients are contained in
$\tilde\ka_{e+}$, $\tilde\ka_{e-}$ and $\tilde\ka_{\rm tr}$,
while all parity-odd coefficients
are in $\tilde\ka_{o+}$ and $\tilde\ka_{o-}$.
The matrix $\tilde\ka_{o+}$ is antisymmetric while
the other three are symmetric.

The form of this decomposition helps 
in determining the portion of the parameter space
to which experiments are sensitive and how different experiments
might overlap.
For example,
typical laboratory experiments with electromagnetic cavities 
search for rotation-violating parity-even observables.
The sensitivity of such experiments is therefore
expected to be dominantly to
the ten rotation-violating parity-even coefficients 
$\tilde\ka_{e+}$ and $\tilde\ka_{e-}$.
For those observables depending
at leading order on the velocity,
the eight coefficients $\tilde\ka_{o+}$ and $\tilde\ka_{o-}$
can be expected to play a role.
Finally,
at second order in the velocity
one can expect the sole rotation-invariant quantity 
$\tilde\ka_{\rm tr}$ to affect measurements.
These considerations are confirmed 
by the results of the detailed analysis
in the sections below.

As another example of the use of the decomposition
\rf{kappas2},
recall that birefringence is known to depend 
on ten linearly independent combinations 
of the components of $\kf$,
which can be chosen as
\cite{km}
\bea
k^a&=&\bigl( 
(\kf)^{0213},~
(\kf)^{0123},~
\nonumber\\
&&\:
(\kf)^{0202}-(\kf)^{1313},~
(\kf)^{0303}-(\kf)^{1212},~
\nonumber\\
&&\:
(\kf)^{0102}+(\kf)^{1323},~
(\kf)^{0103}-(\kf)^{1223},~
\nonumber\\
&&\:
(\kf)^{0203}+(\kf)^{1213},~
(\kf)^{0112}+(\kf)^{0323},~
\nonumber\\
&&\:
(\kf)^{0113}-(\kf)^{0223},~
(\kf)^{0212}-(\kf)^{0313}
\bigr).
\label{ka}
\eea
Relating these to the $\tilde\ka$ matrices,
we find 
\bea
(\tilde\ka_{e+})^{jk} &=&
-\left(
\begin{array}{ccc}
-(k^3+k^4) & k^5 & k^6 \\
k^5 & k^3 & k^7 \\
k^6 & k^7 & k^4 
\end{array}
\right), 
\nonumber \\
(\tilde\ka_{o-})^{jk} &=&
\left(
\begin{array}{ccc}
2k^2 & -k^9 & k^8 \\
-k^9 & -2k^1 & k^{10} \\
k^8 & k^{10} & 2(k^1-k^2) 
\end{array}
\right).
\eea
In this way, 
we can see directly that
birefringence is controlled by the matrices 
$\tilde\ka_{e+}$ and $\tilde\ka_{o-}$.

In terms of the $\ka$ matrices defined in Eq.\ \rf{kappas},
and assuming as before that $(\kaf)^\al = 0$,
the lagrangian \rf{lagrangian} becomes
\bea
\cl&=&\half(\vec E^2-\vec B^2)
+\half \vec E\cdot(\ka_{DE})\cdot\vec E
-\half\vec B\cdot(\ka_{HB})\cdot\vec B
\nonumber\\
&&
+\vec E\cdot(\ka_{DB})\cdot\vec B.
\label{lagkappa}
\eea
Similarly,
using instead the $\tilde \ka$ matrices defined in Eq.\ \rf{kappas2},
we find
\bea
\cl&=&\half[(1+\tilde\ka_{\rm tr})\vec E^2
-(1-\tilde\ka_{\rm tr})\vec B^2]
\nonumber\\
&&
+\half \vec E\cdot(\tilde\ka_{e+}+\tilde\ka_{e-})\cdot\vec E
-\half\vec B\cdot(\tilde\ka_{e+}-\tilde\ka_{e-})\cdot\vec B 
\nonumber\\
&&
+\vec E\cdot(\tilde\ka_{o+}+\tilde\ka_{o-})\cdot\vec B .
\label{lagkappa2}
\eea 
The form of Eq.\ \rf{lagkappa2}
shows that a nonzero coefficient $\tilde\ka_{\rm tr}$
shifts the effective permittivity $\ep$
and effective permeability $\mu$
by $(\ep-1)=-(\mu^{-1}-1)=\tilde\ka_{\rm tr}$, 
corresponding to a shift in the speed of light.
However,
it is possible to remove an overall shift in the
speed of light by making advantageous
coordinate transformations accompanied 
by suitable field redefinitions,
which combine to set $\ep=\mu^{-1}=1$
and transfer the Lorentz violation to a different
sector of the theory.
An explicit example of this procedure is provided
in the next subsection for a toy model involving scalar QED.

In the general context of the standard-model extension,
such transformations modify various
other coefficients for Lorentz violation.
In fact,
similar transformations can move the nine independent coefficients 
$\tilde\ka_{e-}$, $\tilde\ka_{o+}$, and $\tilde\ka_{\rm tr}$
into other sectors of the theory.
Note that this effect is frame dependent
because the coefficients mix under boosts.
Note also that the possibility of absorbing 
$\tilde\ka_{e-}$, $\tilde\ka_{o+}$, $\tilde\ka_{\rm tr}$
elsewhere offers insight as to why birefringence experiments,
which directly compare light with light,
are insensitive to these coefficients.
However,
cavity experiments involve comparisons of radiation with matter,
so all 19 coefficients are observables in this case.

\subsection{Connection to Some Test Models}
\label{others}

Several phenomenological test models 
for Lorentz properties of light have been proposed. 
The standard-model extension
contains all observer-independent sources
of Lorentz violation in terms of known particles,
so it is expected to incorporate 
the existing test models as special cases.
In this subsection,
we comment on the relationships to some popular test models.

Since typical test models assume only one type of matter
other than the photon,
it suffices for our purposes to consider a toy version 
of the standard-model extension that includes only one scalar field
and a limited type of Lorentz violation.
We therefore work with a model 
of Lorentz-violating scalar QED,
defined by the lagrangian 
\bea
\cl&=&
(\et^{\mu\nu}+ (k_\ph)^{\mu\nu})(D_\mu\ph)^\dagger D_\nu\ph
-m^2\ph^\dagger\ph 
\nonumber\\
&&
-\frac14F_{\mu\nu}F^{\mu\nu}
-\frac14\kfi F^{\ka\la}F^{\mu\nu}.
\label{toy}
\eea
In this expression,
the covariant derivative 
takes the usual form,
$D_\mu\ph = \prt_\mu \ph + i q A_\mu \ph$,
and for simplicity we have limited the types of Lorentz violation
to those described by a real symmetric coefficient
$(k_\ph)^{\mu\nu}$ 
and by a coefficient $\kfi$ 
of the type in Eq.\ \rf{lagrangian}.

An interesting test model for Lorentz violation 
is provided by the kinematical framework of Robertson
\cite{hpr}
and its extension to arbitrary synchronizations
by Mansouri and Sexl
\cite{ms}.
These approaches suppose the existence of a ``preferred'' frame 
in which light propagates isotropically
as measured by a standard set of rods and clocks.
The Lorentz transformation between observers
is then generalized to incorporate small changes 
from the conventional boosts in special relativity.
Within a given synchronization,
three parameters $g_0$, $g_1$, $g_2$ are needed 
to fix the generalized Lorentz transformation
and hence to characterize the Lorentz violation.

The construction of the generalized Lorentz transformation
can be illustrated in the context of the model \rf{toy}.
Consider the special case of the model 
for which only the coefficient $(k_\ph)^{00}$ 
is nonzero in a certain frame $\Si$.
Writing this coefficient as
$(k_\ph)^{00}= k^2 -1$,
where $k^2$ deviates slightly from 1, 
the lagrangian takes the form
\bea
\cl&=&
(D_\mu\ph)^\dagger D^\mu\ph
+(k^2 -1) |D_0\ph|^2
-m^2\ph^\dagger\ph 
\nonumber\\ &&
+\half(\vec E^2 - \vec B^2).
\label{toy2}
\eea
In the $\Si$ frame,
the propagation of light is rectilinear and isotropic,
so it may be identified with the preferred frame of the test model.
The Lorentz violation appears only in the $\ph$ sector
of the lagrangian,
which we can suppose describes the detailed physics 
of the rods or clocks in the test model.

The generalized Lorentz transformations ${T^\mu}_\nu$
considered in the kinematical test models 
are the linear transformations $x'^\mu={T^\mu}_\nu x^\nu$ 
from the preferred frame $\Si$
to a coordinate system $S$ attached to an observer
moving at constant velocity in the preferred frame.
By construction,
the observer $S$ defines coordinates 
using the same rods and clocks
and a prescribed synchronization.
However,
in the present context the Lorentz-violating properties 
of the rods and clocks are fixed by the 
Lorentz-violating scalar term in the lagrangian \rf{toy2}.
The generalized Lorentz transformations ${T^\mu}_\nu$
from $\Si$ to $S$ are therefore also determined
in the context of the model \rf{toy}.
They are the transformations leaving invariant the scalar sector
and hence preserving the combination 
$\et^{\mu\nu}+(k_\ph)^{\mu\nu}$
up to a possible resynchronization.
For example,
for the special case of Eq.\ \rf{toy2},
the Robertson parameters are found to be
$g_0 = 1/g_1 = \sqrt{(1- \be^2)/(1-k^2 \be^2)}$,
$g_2 = 1$.
The corresponding Mansouri-Sexl parameters are
$a = 1/b = \sqrt{(1-k^2 \be^2)}$,
$d=1$,
with $\ep = -\be (1- k^2 \be^2)/(1- \be^2)$
in Einstein synchronization
or $\ep = -k^2 \be$ in slow-clock synchronization.
In contrast, 
the standard Lorentz transformations ${\La^\mu}_\nu$
preserve $\et^{\mu\nu}$.

In this simple example, 
the transformation ${T^\mu}_\nu$ leaves invariant the rods and clocks,
while ${\La^\mu}_\nu$ leaves invariant the speed of light.
Both are equally valid.
In the frames related by ${T^\mu}_\nu$,
observers agree on rod lengths and clock rates 
but disagree on the velocity of light.
Moreover, 
the velocity of light is no longer isotropic as measured by
these rods and clocks.
In contrast,
observers related by Lorentz transformations 
agree that light propagates isotropically with speed 1 
but may disagree on rod lengths and clock rates.
The description is a matter of coordinate choice,
and one can move freely from one to the other using 
${T^\mu}_\nu$, ${\La^\mu}_\nu$,
and their inverses.

Note that a ``preferred'' frame in which light propagates isotropically
typically fails to exist in the full standard-model extension,
although in principle one can impose the existence of such a frame
by suitably restricting the coefficients for Lorentz violation.
From this perspective,
the special status enjoyed by photons relative 
to other particles in the kinematical test models
appears somewhat unnatural,
and the structure of the standard-model extension offers 
more general possibilities for kinematical frameworks.
Note also that the standard-model extension 
addresses modifications to all known particles,
so the effects on physical rods and clocks can be directly analyzed.
This is infeasible in kinematical frameworks,
which consider the transformations between frames 
rather than the underlying physics.

Another interesting test model is the $c^2$ model
\cite{hw},
developed for application to studies of Lorentz invariance
as a limiting case of the $TH\ep\mu$ formalism
\cite{ll,cw}.
The $c^2$ model is defined by a lagrangian describing 
the behavior of classical pointlike test particles
in the presence of electromagnetic fields.
The model assumes the existence of a ``preferred'' frame
in which the limiting speed of the test particles is 1,
while the speed of light is $c$.

To see the relation between the $c^2$ model and 
the model \rf{toy},
consider another lagrangian written in a frame $S$ as
\bea
\cl&=&
(D_\mu\ph)^\dagger D^\mu\ph
-m^2\ph^\dagger\ph 
+\half(\vec E^2 - k^2 \vec B^2),
\label{toy3}
\eea
where $k^2$ deviates slightly from 1 as before.
In this theory,
the Lorentz violation appears in the photon sector.
With the identification $k = c$,
the lagrangian for this sector is identical to that of
the $c^2$ model.
Moreover,
the $\ph$ sector is conventional,
representing a quantum field theory 
of minimally coupled scalar particles.
The model \rf{toy3} can therefore be regarded
as the field-theoretic equivalent of the $c^2$ model.

The two models \rf{toy3} and \rf{toy2}
are related by the coordinate transformation
$t \to t/k$, $\vec x \to \vec x$
followed by the field redefinition
$A_\mu \to A_\mu/k$ and charge rescaling $q \to kq$.
They therefore describe the same physics.
Although it is possible to choose coordinates 
so that either the photon or the scalar propagates conventionally, 
the Lorentz violation cannot be 
eliminated simultaneously from both sectors.

We thus see that the $c^2$ model is contained
in the theory \rf{toy} as a special case.
In the terminology of Eq.\ \rf{kappas2},
the parameter $c^2$ could be identified with
the combination of coefficients 
$(1- \tilde\ka_{\rm tr})/(1+ \tilde\ka_{\rm tr})$,
as can be seen from Eq.\ \rf{lagkappa2}.
However,
caution is required in interpreting bounds obtained
with the $c^2$ model in terms of $\tilde\ka_{\rm tr}$
because the identification is valid 
only in a frame $S$ with conventional particles,
which typically fails to exist
in the standard-model extension.

\section{ASTROPHYSICAL TESTS}
\label{bire}

In this section, 
we consider radiation propagating in free space.
The Lorentz-violating electrodynamics predicts birefringence,
which allows sensitive tests of Lorentz symmetry
from observations of radiation propagated over astrophysical distances.
We begin with some general theory,
and then we obtain two sets of bounds on Lorentz violation
from velocity and birefringence constraints.

\subsection{General Theory}
\label{biretheory}

The basic features of plane-wave solutions
to the Lorentz-violating electrodynamics
have been presented in Refs.\ \cite{ck,km},
so only relevant essentials are given here.
With the standard ansatz
$F_\mn (x)= F_\mn (p) e^{-i p_\al x^\al}$
for a plane wave with wave 4-vector $p^\al = (p^0, \vec p)$,
the equation determining the dispersion relation
and the electric field $\vec E$
is the modified Amp\`ere law
\beq
M^{jk}E^k \equiv
(-\de^{jk} p^2 - p^j p^k -2(\kf)^{j \be \ga k} p_\be p_\ga)E^k = 0\ .
\label{ampere}
\eeq
The dispersion relation is obtained as usual
by requiring vanishing determinant of $M^{jk}$.
It suffices for our present purposes to consider
only leading-order effects 
in the coefficients $\kfi$ for Lorentz violation.
To leading order,
one finds
\beq
p^0_\pm=(1+\rh\pm\si)|\vec p|\ ,
\label{dispersion}
\eeq
where
\bea
\rh &=& -\half {\kt_\al}^{\pt{al}\al},
\quad
\si^2 = \half(\kt_{\al\be})^2-\rh^2,
\label{rhsi}
\eea
with
\bea
\kt^{\al\be}&=&(\kf)^{\al\mu\be\nu}{\pht}_\mu {\pht}_\nu,
\quad
{\pht}^\mu = {p^\mu}/{|\vec p|}.
\eea
Note that ${\vec p}^2\rh$ and ${\vec p}^2\si$
are observer Lorentz scalars,
which implies $\rh$ and $\si$
are scalars under observer rotations.

The dispersion relation \rf{dispersion} has two solutions,
with corresponding electric fields $\vec E_\pm$.
In conventional electrodynamics, 
the dispersion relation is $p^0=|\vec p|$
and all fields $\vec E$ perpendicular to $\vec p$ are solutions,
so the propagation is independent of the polarization.
However, 
in the present case
the propagation is goverened by two specific modes
$\vec E_\pm$,
with the general solution to Eq.\ \rf{ampere}
being any linear combination of the two.
This leads to birefringence:
light generically has two components,
each propagating independently.

There are several possible definitions 
for the velocity of the radiation,
including the phase velocity 
$v_p^j \equiv p_0 p^j / {\vec p}^2$,
the group velocity 
$v_g^j \equiv ({\grad_{\vec p})^j}p^0$,
and the velocity of energy transport 
$v_e^j \equiv \th^{j0}/\th^{00}$,
where $\th^{\mu\nu}$ is the energy-momentum tensor.
With the analogy discussed in Sec.\ \ref{defs},
one can show by standard arguments that $\vec v_g = \vec v_e$
for a wave with fixed $\vec p$.
Also, Eq.\ \rf{dispersion} can be used to find
explicit leading-order expressions for the
magnitudes of the phase and group velocities.
We thereby obtain $v_p = v_g = v_e= 1+\rh\pm\si$
to leading order in $\kfi$.
Note also that,
to leading order,
we can write $\pht^\al=(p^0,\vec p)/|\vec p| \approx (1,\hat v)$ 
in the expressions \rf{rhsi} for $\rh$ and $\si$.
The quantity $\hat v$ can be regarded as
the direction of propagation of the radiation,
since the difference between it and the other velocities
arises only at higher order and is irrelevant here.

The mode dependence of the velocity offers interesting
possibilities for experimental tests of the theory.
The velocity difference is 
\beq
\De v \equiv v_+-v_- = 2\si,
\label{dev}
\eeq
and is expected to be tiny.
However, 
for sufficiently large path lengths this
difference might become apparent
in the form of observable effects 
on the pulse shape or the polarization of radiation. 
In the next two subsections,
we exploit these features
to obtain constraints on $\kfi$.

An explicit form for the solutions
$\vec E_\pm$ is needed for some of the analysis.
Using the dispersion relation \rf{dispersion},
the matrix in the Amp\`ere law \rf{ampere}
can be written
\beq
M_\pm^{jk}=
-[2(\rh\pm\si)\de^{jk}+\hat p^j \hat p^k -2\kt^{jk}]\vec p^{\:2} .
\label{Mjk}
\eeq
The form of this matrix shows that 
the solutions $\vec E_\pm$ are wavelength independent
but vary with the direction of propagation.
Also, 
at leading order in $\kfi$
the difference $M_+ - M_-$ is proportional to the identity,
so the leading-order solutions $\vec E_+$
and $\vec E_-$ are perpendicular.
In fact,
at leading order, 
$\vec E_\pm$ are perpendicular
to $\vec p$ as well.

To express $\vec E_\pm$ explicitly, 
a choice of inertial frame must be made.
It is convenient to adopt a standard reference frame 
to report the results of observations
and hence ultimately to place constraints 
on the set of coefficients $\kf$.

A natural choice for the reference frame 
is a Sun-centered celestial equatorial frame 
with the $Z$-axis aligned along the celestial north pole
at equinox 2000.0.
The $Z$-axis is then at a declination of $90^\circ$,
and the $X$- and $Y$-axes lie at declination $0^\circ$ 
and can be chosen to be 
at right ascension $0^\circ$ and $90^\circ$, respectively.
The unit vector $\hat X$ thus points towards the
vernal equinox on the celestial sphere.
The time $T$ is chosen such that
$T=0$ when the Earth crosses the
$XY$ plane on a descending trajectory.
In what follows,
we adopt this standard frame to report results.

For the practical determination of $\vec E_\pm$
for a given wave,
it is easiest first to work 
in a special `primed' frame chosen for that wave.
The result of the calculation can
then be related to the standard Sun-centered frame
by performing a suitable observer rotation.
A convenient primed frame for a given wave is 
the frame in which the wave 4-vector takes the form
$\hat p'^\al=(1;0,0,1)$ to leading order.
The solution for $\vec E_\pm$ can be expressed explicitly 
in terms of the coefficients
$\kf'$ in this frame.
Up to a normalization,
it is found to be
$\vec E_\pm\propto(\sin\xi,\pm1-\cos\xi,0)$,
where
$\tan\xi = 2\kt '^{\, 12}/(\kt '^{\, 11}-\kt '^{\, 22})$.
The two modes are thus linearly polarized.

From the solutions $\vec E_\pm$ and dispersion relation,
it is evident that $\si$ and $\xi$ are the relevant
parameters for birefringent effects for a particular source.
In particular,
$\si \sin\xi = \kt '^{\, 12}$ 
and 
$\si \cos\xi = \half(\kt '^{\, 11}-\kt '^{\, 22})$
represent the minimal linear combinations of $\kf'$ 
that govern birefringence.
The parameter $\rh$ is common to both modes,
but does not contribute to birefringence and
cannot be detected in the experiments discussed below.

The results in the primed frame can be related to the 
standard frame by a suitable observer rotation,
described in Appendix \ref{birefvec}.
The direction of travel of the light in the standard frame
determines two vectors $\vs_s^a$ and $\vs_c^a$ in $k^a$ space
(see Eq.\ \rf{vectors}),
and it turns out that the birefringence of the light 
depends on the two specific linear combinations 
of the coefficients $k^a$ in Eq.\ \rf{ka}
that are parallel to these vectors.

\subsection{Velocity constraints}
\label{velocity}

For the two radiation modes $\vec E_\pm$
propagating over a distance $L$,
the velocity difference \rf{dev}
induces a difference $\De t\approx \De vL$
between the two travel times.
Local measurements made on radiation emitted as a single burst 
from a distant source
can therefore provide sensitivity to the coefficients $k^a$
for Lorentz violation
\cite{astro}.

To apply this idea,
it is useful to consider distant sources 
that produce radiation in a relatively narrow burst 
characterized by a small width $w$,
such as millisecond pulsars or sources of gamma-ray bursts.
These sources typically produce essentially unpolarized radiation,
so the intensity of each mode should be comparable.
The burst can then be regarded as
a superposition of two independently propagating pulses,
one for each mode.
For a sufficiently great distance $L$,
a nonzero $\De v$ would cause the two pulses to separate
enough to become distinguishable.
This type of signal would manifest itself
as two pulses with similar time structure
but differing in arrival time.
The pulses would each be linearly polarized,
and they would have mutually perpendicular polarization angles.

If only a single pulse is observed,
a limit on Lorentz violation can be deduced.
The relationship between the
observed pulse width $w_o$ and the source pulse width $w_s$ 
is approximately $w_o \approx w_s + \De t$.
Observations of $w_o$ 
can therefore be used to obtain a conservative
bound on 
$\De t = \De v L = 2 \si L$
and hence a bound on the coefficients $k^a$.

\begin{center}
\begin{tabular}{|l||rl|rl|c|}
\hline
\multicolumn{1}{|c||}{Source} &
\multicolumn{2}{c|}{$L$} &
\multicolumn{2}{c|}{$w_o$} & Ref.\ \\ 
\hline \hline
GRB\,971214     & 2.2 &Gpc& 50  &s&      \cite{kulk2,batse} \\
GRB\,990123     & 1.9 &Gpc& 100 &s&      \cite{batse,kulk1} \\
GRB\,980329     & 2.3 &Gpc& 50  &s&      \cite{batse,frutcher} \\
GRB\,990510     & 1.9 &Gpc& 100 &s&      \cite{batse,gcn324} \\
GRB\,000301C    & 2.0 &Gpc& 10  &s&      \cite{gcn568,gcn605} \\
PSR\,J1959+2048 & 1.5 &kpc& 64  &$\mu$s& \cite{psrcat} \\
PSR\,J1939+2134 & 3.6 &kpc& 190 &$\mu$s& \cite{psrcat} \\
PSR\,J1824-2452 & 5.5 &kpc& 300 &$\mu$s& \cite{psrcat} \\
PSR\,J2129+1210E& 10.0 &kpc& 1.4 &ms&     \cite{psrcat} \\
PSR\,J1748-2446A& 7.1 &kpc& 1.3 &ms&     \cite{psrcat} \\
PSR\,J1312+1810 & 19.0 &kpc& 4.4 &ms&     \cite{psrcat} \\
PSR\,J0613-0200 & 2.2 &kpc& 1.4 &ms&     \cite{psrcat} \\
PSR\,J1045-4509 & 3.2 &kpc& 2.2 &ms&     \cite{psrcat} \\
PSR\,J0534+2200 & 2.0 &kpc& 10  &$\mu$s& \cite{psrcat,gpp2} \\
PSR\,J1939+2134 & 3.6 &kpc& 5   &$\mu$s& \cite{psrcat,gpp1} \\
\hline
\end{tabular}
\end{center}
\begin{center}
Table 1. Source data for velocity constraints. 
\end{center}

Table 1 lists data for fifteen sources
suitable for placing this type of constraint.
The first five lines list 
gamma-ray bursts with known redshifts.
The widths listed for these contain 
all significant time structure of the pulse.
The distance $L$ is determined from the redshift
by the look-back time in a conservative cosmology 
for a matter-dominated universe with Hubble constant
$H_0= 80$ km s$^{-1}$ Mpc$^{-1}$.
The next eight sources in the table are millisecond pulsars.
The listed pulse width is that at $10 \%$ peak intensity.
The final two sources are giant-pulse pulsars.
These exhibit intense pulses with
characteristic widths on the order of several $\mu$s.

For each source in Table 1, we take
$\si \leq w_o/2 L$ 
as a bound on $\si$.
For a single source, 
this places constraints on a
two-dimensional subspace of the full 10-dimensional
parameter space of the coefficients $k^a$.
The subspace is represented by the linear combinations
$\si\sin\xi$ and $\si\cos\xi$ associated with that
particular source.
To bound all ten coefficients $k^a$,
ten linearly independent constraints of this type are needed.
This is feasible using five or more
sources at different positions on the sky.

We proceed by assuming the constraint for each source
in Table 1 is consistent with a measurement of $\si=0$,
and we take the bound $\si \leq w_o/2 L$ 
as a reasonable estimate of the error in a null measurement.
The associated $\ch^2$ distribution is
$\ch^2=\sum 4L^2\si^2/w_o^2$,
where the sum is over the fifteen sources.
This is a quadratic form in $k^a$.
Considering $|k^a|$ and minimizing $\chi^2$ with respect
to the other nine degrees of freedom, 
we obtain a bound of 
\beq
|k^a| < 3\times10^{-16}
\label{velbound}
\eeq
in the Sun-centered celestial equatorial frame,
at the $90\%$ confidence level.

This bound is much less stringent than that obtained
through polarization measurements,
as discussed below.
However, 
the method is relatively straightforward
and avoids some of the complexities involved 
in the polarization analysis.

\subsection{Polarization constraints}
\label{polar}

In this subsection, we expand on the material found
in Ref.\ \cite{km}.
An improvement on the previous result is made by considering
the cosmological redshift of light.

A general electric field $\vec E$ can be decomposed
into its birefringent components $\vec E_\pm$.
Defining unit vectors
$\hat\ve\pm=\vec E_\pm/|\vec E_\pm|$,
the decomposition is 
\beq
\vec E(x) = (E_+ \hat\ve_+e^{-ip^0_+t}+
E_- \hat\ve_-e^{-ip^0_-t})e^{i\vec p \cdot\vec x}.
\eeq
The differing phase velocities of the two modes 
results in a change in relative phase as the wave propagates,
given by \cite{km}
\beq
\De\ph =( p^0_+-p^0_-)t 
\approx 2\pi \De v_p L/\la \approx 4\pi\si L/\la,
\label{deph}
\eeq
where $\De v_p$ is the difference in phase velocities, 
$\la$ is the wavelength,
and $L$ is the distance traveled.
The phase change modifies the polarization state of the radiation,
with larger effect for more distant sources.
Appendix \ref{polbackground} provides 
a brief review of pertinent concepts involving polarization
in the present context.

\begin{figure}[]
\centerline{
\psfig{figure=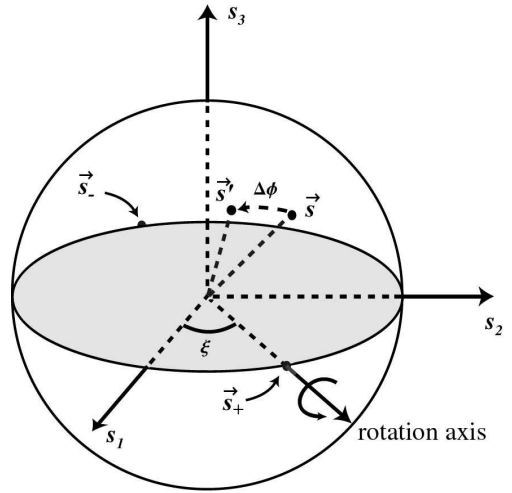,width=0.8\hsize}}
\caption{Rotation of the Stokes vector
about $\vec s_+=-\vec s_-$.}
\label{sphere2}
\end{figure}

In the primed frame described in Sec.\ \ref{biretheory},
the Stokes vectors for $\hat\ep_\pm$ are 
$\vec s_\pm = \pm(\cos\xi,\sin\xi,0)$.
These vectors correspond to opposite points on
the equator of the Poincar\'e sphere,
as expected for linearly polarized modes.
As described in Appendix \ref{polbackground},
the axis of rotation induced by the phase change
$\De\ph=4\pi\si L/\la$
is therefore in the $s^1$-$s^2$ plane.
This affects both $\psi$ and $\chi$,
as can be seen from Fig.\ \ref{sphere2}.

A change in phase can arise from a change
in either $L$ or $\la$.
The induced change in the polarization
depends not only on $\kf$ 
but also on the initial polarization.
For cosmological sources, 
it may be impossible to determine independently 
the polarization at the source,
in which case one cannot determine 
whether a change in polarization is strictly 
due to a change in $L$.
It is therefore of more interest to
focus on the wavelength dependence of the polarization change.
Making the reasonable assumption
that the emitted polarization is relatively constant
over a given range of wavelengths,
the relevant quantity becomes the phase shift
as a function of wavelength,
\beq
\de\ph=4\pi\si L \bigr(\fr 1 {\la} - \fr 1 {\la_0}\bigl),
\label{dephref}
\eeq
relative to a reference wavelength $\la_0$.
Standard spectropolarimetric techniques then
allow a measurement of this effect.
Note that knowledge of physical processes
in certain classes of objects producing the polarized radiation 
might make it feasible to include a known initial polarization
in the analysis,
but this is unlikely to improve significantly the constraint 
obtained here.

The effect on the measured polarization 
as the wavelength is changed
can be visualized using the Poincar\'e sphere.
Suppose a source produces radiation with constant
polarization over a range of wavelengths.
This radiation can be represented by a single point
on the Poincar\'e sphere.
As the light propagates towards the Earth,
the presence of Lorentz violation causes this point to rotate 
along an arc on the sphere.
For any fixed wavelength,
the rotation axis and rate depend on the coefficients $\kfi$
and on the position of the source on the sky.
However, Eq.\ \rf{dephref} shows 
that shorter wavelengths rotate more than longer ones.
Therefore,
as measurements of the Stokes vector
are made over a range of wavelengths,
the results trace a circular arc 
on the surface of the Poincar\'e sphere.

Let $\ps_0$ and $\ch_0$ represent the observed
polarization of a point on this arc with the
reference wavelength $\la_0$.
Using this point as a reference, 
we require the change in polarization relative to this point 
induced by Eq.\ \rf{dephref}.
This polarization change is given by
\beq 
s^j(\ps,\ch)=m^{jk}(\de\ph)s^k(\ps_0,\ch_0),
\eeq
where $m^{jk}$ is the rotation matrix about $\vec s_+$ by $\de\ph$.
The matrix $m^{jk}$ is analogous 
to the Mueller matrix used in polarimetry
to describe the effects of various filters and polarizers on light.
Its explicit form is given as Eq.\ (8) of Ref.\ \cite{km}.
The angle $\ch$,
which controls the amount of circular polarization,
is absent from most published spectropolarimetric data.
It is therefore most effective to focus attention
on the change $\de\ps=\ps-\ps_0$ in $\ps$ 
from the reference value $\ps_0$,
which is given as Eq.\ (9) of Ref.\ \cite{km}.
A procedure for fitting this equation to existing
spectropolarimetric data is also provided in this reference,
and a 90\% confidence-level bound 
of $|k^a|< 3\times 10^{-32}$ is obtained from 
spectropolarimetric data for 16 cosmological sources.

In the present work,
we use the same procedure to obtain a slight improvement
on the existing bound
by incorporating the redshift of the light
as it propagates to the Earth.
Cosmological redshift implies that over the path traveled 
the light has shorter wavelength than observed.
Taking the same conservative cosmology as in the
previous subsection and integrating the phase change
over the propagation time yields 
\beq
\De\phi = \fr {4\pi\si}{\la}\fr 2{H_0}
\bigl(1-\fr 1 {\sqrt{1+z}}\bigr),
\eeq
where $\la$ is the observed wavelength and $z$ is the redshift.
To account for the redshift,
it therefore suffices to replace $L$ with 
$L_{\rm eff}=2(1-(1+z)^{-1/2})/H_0$ 
in the analysis.

\begin{center}
\begin{tabular}{|l||c|c|c|}
\hline
\multicolumn{1}{|c||}{Source} 
& $L_{eff}$~(Gpc) & $10^{30}L_{eff}/\la$ & $\log_{10}\si$ \\ 
\hline \hline
IC 5063 \cite{hough}                      & 0.04 & 0.56 - 2.8 & -30.8 \\
3A 0557-383 \cite{brindle}                & 0.12 & 2.2 - 8.5  & -31.2 \\
IRAS 18325-5925 \cite{brindle}            & 0.07 & 1.0 - 4.9  & -31.0 \\
IRAS 19580-1818 \cite{brindle}            & 0.14 & 1.8 - 9.3  & -31.0 \\
3C 324 \cite{cimatti465}                  & 2.44 & 82 - 180   & -32.3 \\
3C 256 \cite{dey}                         & 3.04 & 110 - 220  & -32.4 \\
3C 356 \cite{cimatti476}                  & 2.30 & 78 - 170   & -32.3 \\
F J084044.5+\ldots \cite{brothertonfirst} & 2.49 & 88 - 170   & -32.4 \\
F J155633.8+\ldots \cite{brothertonfirst} & 2.75 & 99 - 160   & -32.4 \\
3CR 68.1 \cite{brotherton}                & 2.48 & 84 - 180   & -32.4 \\
QSO J2359-1241 \cite{brothertonqso}       & 2.01 & 110 - 120  & -31.2 \\
3C 234 \cite{kishimoto}                   & 0.61 & 55 - 81    & -31.7 \\
4C 40.36 \cite{vernet}                    & 3.35 & 120 - 260  & -32.4 \\
4C 48.48 \cite{vernet}                    & 3.40 & 120 - 260  & -32.4 \\
IAU 0211-122 \cite{vernet}                & 3.40 & 120 - 260  & -32.4 \\
IAU 0828+193 \cite{vernet}                & 3.53 & 130 - 270  & -32.4 \\
\hline
\end{tabular}
\end{center}
\begin{center}
Table 2. Source data for polarization constraints.
\end{center}

Table 2 lists 16 sources with published values of $\ps$.
The second column of the table displays the effective
distance $L_{\rm eff}$ traveled by the light.
The third column provides the range of wavelengths
for which data are used.
In fitting to $\de\ps$, 
we choose to set $\ps_0$ equal to the mean value
of the measured $\ps$.
For each source, 
$\ch_0$ and $\la_0$ are fitted to the data.
These angles can be thought of as the two degrees of freedom
needed to describe the unknown polarization at the source.
Ideally, 
at this point the data would be fitted to
all the sources simultaneously.
However, 
since $k^a$ has 10 elements and each source
introduces two additional parameters, 
this would be involved.
Instead,
we examine each source individually and
look for the desired wavelength dependence.

Adopting the same analysis strategy as described
in Ref.\ \cite{km}
yields the bounds for each source listed in the 
last column of Table 2,
which can be combined to yield a bound of
\beq
|k^a|<2 \times 10^{-32}
\label{polbound}
\eeq
in the Sun-centered celestial equatorial frame,
at the $90\%$ confidence level.

\section{LABORATORY TESTS}
\label{labexp}

The Lorentz-violating electrodynamics predicts 
shifts in cavity-resonance frequencies,
which offers the opportunity for  
sensitive tests of Lorentz symmetry 
in laboratories on the Earth and in space.
This section presents a general framework for
the analysis of such experiments.
We begin with some general considerations
and then separately consider in more detail
the cases of optical cavities and microwave cavities.

\subsection{General Considerations}
\label{labtheory}

Many tests of special relativity search 
for variations in some observable
that might arise from the rotation or boost of the apparatus
due to the motion of the Earth.
Lorentz-violating theories predict periodic variations 
at multiples of the Earth's sidereal or orbital frequencies.
For example,
high-sensitivity measurements of coefficients in  
the fermion sector of the standard-model extension 
have been performed by comparing two clocks as the Earth rotates 
\cite{ccexpt}.
The clocks are typically the frequencies 
associated with specific Zeeman atomic transitions,
and the standard-model extension predicts variations
in these frequencies with the orientation of the apparatus
and hence with the Earth's rotation.
Similar tests could be performed in space,
with the frequency variations depending on
the orbital and rotational properties of the spacecraft
\cite{bklr}.

Resonant cavities can also serve as clocks,
and they can be used in clock-comparison experiments
to test properties of electromagnetic fields 
instead of atomic transitions.
In particular,
clock-comparison experiments of this type
can be used to probe the photon sector
of the standard-model extension.
One relevant issue in the analysis of these experiments
is establishing the transformation 
between the laboratory frame and a standard celestial frame.
Another is the determination of the predicted frequency shifts. 
In this subsection,
these issues are addressed in a general context.

\subsubsection{Generic laboratory experiment}

Consider a general laboratory-based experiment
measuring some electrodynamic observable ${\cal O}$.
Typically, 
the constitutive relations \rf{DH}
change the observable from its conventional value ${\cal O}_0$.
We consider a change $\de{\cal O}$,
taken to be linear in the matrices $\ka_{DE}$,
$\ka_{HB}$, and $\ka_{DB}=-\ka_{HE}^T$.
In a frame fixed to the laboratory,
$\de{\cal O}$ can be written as 
\bea
\de{\cal O} &=& ({\cal M}_{DE})^{jk}_{\rm lab}(\ka_{DE})^{jk}_{\rm lab}
+({\cal M}_{HB})^{jk}_{\rm lab}(\ka_{HB})^{jk}_{\rm lab} 
\nonumber \\
&&
+({\cal M}_{DB})^{jk}_{\rm lab}(\ka_{DB})^{jk}_{\rm lab} ,
\label{dOlab}
\eea
where 
$({\cal M}_{DE})_{\rm lab}$, $({\cal M}_{HB})_{\rm lab}$,
and $({\cal M}_{DB})_{\rm lab}$ 
are experiment-specific constant matrices
determined by the apparatus.
The symmetries of the $\ka$ matrices
can be imposed on their ${\cal M}$ counterparts 
when convenient.

Due to the orbital and rotational motion of the Earth
or the space platform,
the laboratory cannot be considered an inertial frame.
As a result, 
the laboratory-frame coefficients
$(\ka_{DE})^{jk}_{\rm lab}$, $(\ka_{HB})^{jk}_{\rm lab}$,
and $(\ka_{DB})^{jk}_{\rm lab}$ 
vary in time.
We can exploit the induced variation in ${\cal O}$
by searching for periodic fluctuations in ${\cal O}$
at the relevant frequencies.
A measurement of this type of variation would be
a signal for Lorentz violation.

To determine the dependence of the 
periodic variation on the coefficients $\kfi$, 
we seek an expression similar to Eq.\ \rf{dOlab}
in an inertial frame.
A suitable choice for a standard inertial frame
is the Sun-centered celestial equatorial frame
defined in Sec.\ \ref{biretheory}.
The coefficients for Lorentz violation in this frame,
$(\ka_{DE})^{JK}$, $(\ka_{HB})^{JK}$,
and $(\ka_{DB})^{JK}$, are constant.

The observer Lorentz transformation between the two frames
can be used to relate the corresponding
two sets of $\ka$ matrices.
Since the velocity of the Earth with respect to the Sun is 
$\be_\oplus \approx 10^{-4}$,
it suffices for our purposes 
to construct the transformation to leading order.
At this order,
the Lorentz matrix $\La^\mu_{\pt{\mu}\nu}$ 
implementing the transformation 
from the Sun-centered frame to the laboratory frame is
\beq
\La^0_{\pt{0} T} = 1 , 
\ \  
\La^0_{\pt{0}J} = -\be^J,
\ \  
\La^j_{\pt{j}T} = -(R\cdot\vec\be)^j, 
\ \  
\La^j_{\pt{j}J} = R^{jJ} ,
\label{lambda}
\eeq
where $\vec\be$ is the velocity of the laboratory with
respect to the Sun-centered frame and $R^{jJ}$ is the
spatial rotation from the Sun-centered frame to the laboratory frame.
Some calculation shows that the
induced transformation between the $\ka$ matrices is 
\bea
(\ka_{DE})_{\rm lab}^{jk}
&=&T_0^{jkJK}(\ka_{DE})^{JK}-T_1^{(jk)JK}(\ka_{DB})^{JK},
\nonumber\\
(\ka_{HB})_{\rm lab}^{jk}
&=&T_0^{jkJK}(\ka_{HB})^{JK}-T_1^{(jk)KJ}(\ka_{DB})^{JK},
\nonumber\\
(\ka_{DB})_{\rm lab}^{jk}&=&T_0^{jkJK}(\ka_{DB})^{JK}
\nonumber\\ &&
+T_1^{kjJK}(\ka_{DE})^{JK} +T_1^{jkJK}(\ka_{HB})^{JK},
\label{trans}
\eea
where
\beq
T_0^{jkJK}=R^{jJ}R^{kK},
\quad
T_1^{jkJK}= R^{jP}R^{kJ} \ep^{KPQ} \be^Q .
\label{Ts}
\eeq
The tensor $T_0$ is a rotation,
while $T_1$ is a leading-order boost contribution.
Although the contributions involving $T_1$ are suppressed by $\be$,
they access distinct combinations of coefficients 
and can introduce different time dependence, 
which may lead to fundamentally different tests.

To apply Eqs.\ \rf{trans} and \rf{Ts},
the laboratory frame must be specified.
Appendix \ref{frames} defines
our standard Earth-based and space-based frames
and establishes the transformations from these to
the reference Sun-centered celestial equatorial frame.

\subsubsection{Cavity experiments}

Two classes of cavities are of interest in the present context:
optical cavities, 
for which the wavelength of the light 
is much smaller than the cavity size,
and microwave cavities,
for which the wavelength and cavity size are comparable.
In both cases, 
the interesting quantity 
is the fractional resonant-frequency shift 
$\de\nu/\nu$.

For a given cavity,
let 
$\vec E_0$, $\vec B_0$, $\vec D_0$, $\vec H_0$ 
be the fields associated with a conventional mode 
of resonant angular frequency $\om_0$.
Nonzero $\kf$ coefficients can perturb these resonance fields.
Let $\vec E$, $\vec B$, $\vec D$, $\vec H$
be the perturbed fields for the resonant mode 
in the presence of Lorentz violation,
and let $\de\nu = \de\om /2\pi$ 
represent the change in the resonant frequency
relative to the conventional case.
A manipulation of the Lorentz-violating Maxwell equations
then yields the fractional resonant-frequency shift as
\bea
\fr{\de\nu}\nu&=&
-\left(\int_V d^3x \bigl(\vec E_0^*\cdot\vec D
+\vec H_0^*\cdot\vec B\bigr)\right)^{-1}
\nonumber\\
&&
\times\int_V d^3x \left(\vec E_0^*\cdot\vec D-\vec D_0^*\cdot\vec E
-\vec B_0^*\cdot\vec H+\vec H_0^*\cdot\vec B\right. 
\nonumber\\
&&\left.
\qquad
\qquad
-i\om_0^{-1}\vec\nabla\cdot
(\vec H_0^*\times\vec E-\vec E_0^*\times\vec H)
\right) ,
\label{dnu}
\eea
where the integrals are over the volume $V$
of the cavity.
This equation holds for any harmonic system,
even for large differences 
between the conventional and perturbed modes.
Note that the divergence term results in a surface integral
over the boundary of $V$.

For the application to Lorentz violation,
the perturbed modes are expected to differ 
only slightly from the unperturbed ones.
Also,
the boundary conditions can reasonably be taken
such that the divergence term in Eq.\ \rf{dnu} vanishes.
The point is that, for leading-order effects,
we can approximate the cavity as lossless
and idealize the surface of the cavity as a perfect conductor.
The boundary condition 
of vanishing surface tangential electric field $\vec E_0$
follows as usual from the Faraday equation 
$\curl E+\prt_0 {\vec B}=0$
and the vanishing of $\vec E_0$ inside the conductor.
The latter can be regarded as a consequence of the Lorentz force law.
To determine the tangential perturbed field $\vec E$ on the 
cavity surface,
we note that Lorentz violation in the photon sector
leaves the force law unaffected.
Disregarding for simplicity any effects on the force law
arising from Lorentz violation in the fermion sector 
of the standard-model extension,
which in any case would be expected to enhance a signal,
it follows that
the tangential component of $\vec E$ also vanishes on the surface.
With these boundary conditions,
the normal component of
$(\vec H_0^*\times\vec E-\vec E_0^*\times\vec H)$
is zero at the surface of the cavity.

For leading-order effects of Lorentz violation,
it suffices to expand the remaining terms of Eq.\ \rf{dnu} 
in the coefficients $\kfi$.
If the cavity is void of matter,
then $\vec D_0=\vec E_0$, $\vec H_0=\vec B_0$,
and the constitutive relations \rf{DH} 
yield the approximate relations
\bea
\vec D-\vec E &\simeq&\ka_{DE}\cdot\vec E_0+\ka_{DB}\cdot\vec B_0 ,
\nonumber \\
\vec H-\vec B &\simeq&\ka_{HE}\cdot\vec E_0+\ka_{HB}\cdot\vec B_0 .
\eea
If the cavity contains matter,
we adopt instead a general linear relation between 
the unperturbed fields $(D_0,H_0)$ and $(E_0,B_0)$ 
and assume for simplicity a lossless medium. 
In either case,
we find that the leading-order fractional frequency shift 
becomes 
\bea
\fr{\de\nu}\nu 
&=& -\fr1{4\expect U}\int_V d^3x \,
\left(\vec E_0^* \cdot \ka_{DE} \cdot \vec E_0
-\vec B_0^* \cdot \ka_{HB} \cdot \vec B_0 \right.
\nonumber \\
&&
\left.
\qquad\qquad\qquad\qquad
+2\Re\bigl(\vec E_0^*\cdot\ka_{DB}\cdot\vec B_0\bigr)\,\right) ,
\label{dnu1}
\eea
where 
$\expect U=\int_V d^3x\,
(\vec E_0\cdot\vec D_0^*+\vec B_0\cdot\vec H_0^*)/4$
is the time-averaged energy stored in the unperturbed cavity.
Note that $\de\nu/\nu$ is real,
reinforcing the argument that the vacuum is lossless
and indicating that the $Q$ factor of the cavity
remains unaffected by Lorentz violation at leading order.

\subsection{Optical cavity experiments}
\label{optcav}

Among the classic tests of Lorentz invariance 
are the Michelson-Morley \cite{mm}
and Kennedy-Thorndike \cite{kt} experiments.
Both concern the speed of light,
with the former searching for spatial anisotropy 
and the latter seeking 
dependence on the laboratory velocity.
The standard-model extension can be used
as a general framework for analyzing these experiments.
In this section,
we consider modern versions of these tests 
that use optical cavities 
to achieve improved sensitivities
\cite{bh,hh,brax}.

\subsubsection{Theory}
\label{optcavth}

We can use the results in Sec.\ \ref{labtheory} to obtain
an expression for the fractional frequency shift
$\de\nu/\nu$
arising from Lorentz-violating effects
in an optical cavity.
The idea is to regard the cavity 
as two parallel reflecting planar surfaces 
with plane waves propagating between them
normal to the surfaces,
and then to apply Eq.\ \rf{dnu1}.

The resonant modes of optical cavities
can be regarded as standing waves.
For simplicity and definiteness,
we suppose the unperturbed cavity contains a medium
having transverse relative permittivity $\ep$ 
and relative permeability $\mu=1$,
with the case $\ep = 1$ corresponding to a cavity void of matter.
As usual,
the unperturbed fields can be taken as
\bea
\vec E_0(x)&=&\vec E_0
\cos(\om_0\hat N\cdot\vec x+\ph)e^{-i\om_0t} ,
\nonumber\\
\vec B_0(x)&=&i\sqrt{\ep}\hat N\times\vec E_0
\sin(\om_0\sqrt{\ep}\hat N\cdot\vec x+\ph)e^{-i\om_0t} ,
\label{reswave}
\eea
where $\hat N$ is a unit vector pointing along the
length of the cavity, 
$\ph$ is a phase,
and $\vec E_0$ is a vector perpendicular to $\hat N$ 
that specifies the polarization.
The conventional resonant frequencies are
given by $\om_0=\pi m / \sqrt{\ep}l$,
where $m$ is an integer
and $l$ is the length of the cavity.

Substitution of Eq.\ \rf{reswave} into
Eq.\ \rf{dnu1} yields the desired result
for the fractional frequency shift:
\bea
\fr{\de\nu}{\nu}
&=&-\fr1 {2|\vec E_0|^2}
\bigl[\vec E_0^*\cdot(\ka_{DE})_{\rm lab}\cdot\vec E_0/\ep 
\nonumber \\
&&
\qquad
-(\hat N \times \vec E_0^*)
\cdot(\ka_{HB})_{\rm lab}\cdot(\hat N \times \vec E_0)\bigr] .
\label{dnuopt}
\eea
This expression for the fractional frequency shift
is also obtained in an alternative approach 
from a different physical perspective,
as described in Appendix \ref{optical}.

The laboratory-frame matrices ${\cal M}_{\rm lab}$ 
introduced in Eq.\ \rf{dOlab}
can be extracted from Eq.\ \rf{dnuopt}.
We find
\bea
({\cal M}_{DE})_{\rm lab}^{jk} &=& 
-\fr{\Re(E_0^*)^j(E_0)^k}{2\ep |\vec E_0|^2} , 
\nonumber \\
({\cal M}_{HB})_{\rm lab}^{jk} &=& 
\fr{\Re(\hat N\times\vec E_0^*)^j
(\hat N\times\vec E_0)^k}{2|\vec E_0|^2} , 
\nonumber \\
({\cal M}_{DB})_{\rm lab}^{jk} &=& 0 \ .
\label{opticalMs}
\eea
These equations show that in the presence of Lorentz violation
the frequency of an optical-cavity oscillator depends 
both on the orientation of the cavity
and on the polarization of the light
with respect to the laboratory frame.

To analyze an experiment with an optical cavity,
one can now proceed as follows.
First,
determine the laboratory-frame matrices $\cal{M}_{\rm lab}$ 
from the apparatus by applying Eq.\ \rf{opticalMs}.
These matrices are constant
if the cavity is fixed in the laboratory 
but vary with time
if the cavity is rotated in the laboratory.
Next, 
relate the laboratory-frame matrices $\ka_{\rm lab}$ 
to those in the reference Sun-centered frame 
using the transformation \rf{trans} and the material
in Appendix \ref{frames}.
The time dependence of the cavity resonant frequency 
can then be calculated
using Eq.\ \rf{dnuopt} or equivalently Eq.\ \rf{dOlab}.
Finally, 
the amplitudes and phases of particular harmonics
can be obtained and compared to the experimental data.

As an illustration of the analysis procedure, 
consider laser light incident on a cavity positioned horizontally 
in an Earth-based laboratory,
with the light linearly polarized along the $z$ axis.
Denote by $\th$ the angle 
between the $x$ axis and the cavity orientation.
Then,
$\hat N = (\cos{\th},\sin{\th},0)$,
and in the laboratory frame
the fractional frequency shift becomes
\bea
\fr {\de\nu}\nu &=&
-\frac14[ 2(\ka_{DE})_{\rm lab}^{33}/\ep
-(\ka_{HB})_{\rm lab}^{11}
-(\ka_{HB})_{\rm lab}^{22} ]
\nonumber \\
&&-\half(\ka_{HB})_{\rm lab}^{12}\sin2\th 
\nonumber \\
&&-\frac14 [(\ka_{HB})_{\rm lab}^{11}-(\ka_{HB})_{\rm lab}^{22}]
\cos2\th  .
\eea

The next step is to transform this result
to the Sun-centered celestial equatorial frame.
Using Eqs.\ \rf{trans} and \rf{Ts},
the fractional frequency shift takes the form
\bea
\fr{\de\nu}\nu &=& A + B\sin2{\th} +C\cos2{\th} , 
\label{ffs}
\eea
where
\bea
A &=& A_0+A_1\sin\om_\oplus T_\oplus+A_2\cos\om_\oplus T_\oplus 
\nonumber \\
&& +A_3\sin2\om_\oplus T_\oplus+A_4\cos2\om_\oplus T_\oplus  , 
\nonumber \\
B &=& B_0+B_1\sin\om_\oplus T_\oplus+B_2\cos\om_\oplus T_\oplus 
\nonumber \\
&& +B_3\sin2\om_\oplus T_\oplus+B_4\cos2\om_\oplus T_\oplus , 
\nonumber \\
C &=& C_0+C_1\sin\om_\oplus T_\oplus+C_2\cos\om_\oplus T_\oplus 
\nonumber \\
&& +C_3\sin2\om_\oplus T_\oplus+C_4\cos2\om_\oplus T_\oplus .
\label{dnu11}
\eea
The quantities 
$A_{0,1,2,3,4}$, $B_{0,1,2,3,4}$, and $C_{0,1,2,3,4}$ 
are linear in the coefficients for Lorentz violation
and depend on the colatitude $\ch$.
They are given explicitly to order $\be$ in Appendix \ref{signalcoeff}.
Note that the coefficient $\tilde\ka_{\rm tr}$ appears
only in $A_0$,
resulting in a constant frequency shift.
It follows that sensitivity to $\tilde\ka_{\rm tr}$ 
is suppressed by at least two powers of $\be$
in this experiment.

The analysis could now proceed along several lines.
One possibility is to adopt the birefringent constraints 
\rf{polbound}.
The expressions in Appendix \ref{signalcoeff} 
can then be simplified by setting 
$(\tilde\ka_{e+})^{JK}=(\tilde\ka_{o-})^{JK}=0$.
This shows that the eight coefficients 
$\tilde\ka_{e-}$, $\tilde\ka_{o+}$ 
are directly accessible through fitting
the measured frequency shift 
to Eq.\ \rf{dnu11}.
Another possibility is to  
include all coefficients in the analysis.
This would provide a direct laboratory check on the
birefringence results.
Although in practice the sensitivity is much reduced,
the systematics of laboratory-based experiments are
fundamentally different from those in cosmological tests
and so this check is worthwhile.
We remark that the isolation of specific coefficients
for Lorentz violation can be aided
by considering different experimental configurations.
These include adopting a different polarization
and rotating the apparatus in the laboratory,
which produces a time dependence in $\th$.

\subsubsection{Experiment}

A modern Michelson-Morley experiment with sensitivity 
to a fractional frequency shift $\de\nu/\nu$
of about $10^{-13}$
was performed by Brillet and Hall 
\cite{bh}.
A similar sensitivity was achieved by Hils and Hall 
\cite{hh}
in a Kennedy-Thorndike experiment,
recently repeated using a cryogenically cooled cavity
by Braxmaier {\it et al.} 
\cite{brax}. 
These experiments compare the fractional frequency shifts
between two lasers.
One laser is stabilized to a molecular transition
and serves as a reference frequency.
A portion of the light from the second laser
is sent into one end of an optical cavity,
and the light emerging at the other end 
is used to tune this laser to the cavity resonant frequency.
The remaining light from the second laser 
is combined with the light from the reference laser,
and the beat frequency is measured.
In the classic analysis, 
the frequency of the reference laser is independent
of violations of special relativity 
while the frequency of the cavity-stabilized laser depends 
on the speed of light along the length of the cavity.

The Brillet-Hall experiment studies spatial isotropy
by seeking changes in the beat frequency
as the cavity is rotated in the laboratory
with a period of about 10 s.
The vector Fourier amplitude 
is measured at twice the cavity rotation frequency.
In the present context,
if we suppose for definiteness a vertical laser polarization
as in the previous subsection,
this experiment offers sensitivity to the quantities
$B$ and $C$ in Eq.\ \rf{ffs}
through the time dependence of $\th$.
The reported fractional frequency shift 
is $1.5\pm 2.5 \times 10^{-15}$.
The analysis yielding this bound supposes a signal at $2\om_\oplus$
and averages over several days of data.
Within the framework leading to Eq.\ \rf{dnu11},
this bound would translate to a constraint on a 
particular combination of the coefficients 
$B_3$, $B_4$, $C_3$, $C_4$
at the level of about a part in $10^{15}$. 

A complete dataset of the type taken in this experiment
could be analysed using Eq.\ \rf{dnu11}
to extract several different measurements of  
combinations of $B_n$ and $C_n$.
For example,
consider the one-day dataset displayed in Fig.\ 2 of Ref.\ \cite{bh}.
In this dataset,
no variation is seen in the frequency above the level of
$\sqrt{B^2+C^2} \lsim 4 \times 10^{-13}$.
In fact, 
these data exhibit a one-day signal
involving a roughly constant Fourier amplitude 
of about $2 \times 10^{-13}$ with nearly constant phase,
attributed to a slight tilt in the rotation platform.
Since a nonzero value of 
$\sqrt{B_0^2+C_0^2}$ would produce a similar signal,
compelling measurements of $B_0$, $C_0$
via this method appear problematic.
However,
bounds on combinations of the quantitites $B_n$, $C_n$ 
for $n\neq 0$
could be extracted by studying the behavior of the data 
at both the sidereal frequency $\om_\oplus$ and
its harmonic $2\om_\oplus$.
As can be seen from the expressions in Appendix \ref{signalcoeff},
the quantities $B_n$, $C_n$ involve unsuppressed combinations of  
the coefficients $\tilde\ka_{e+}$, $\tilde\ka_{e-}$
for Lorentz violation,
along with combinations of the coefficients
$\tilde\ka_{o+}$, $\tilde\ka_{o-}$
suppressed by one power of the velocity. 
It therefore appears feasible to perform a systematic analysis 
of a complete dataset in a Michelson-Morley experiment 
with an optical cavity 
to measure combinations of the coefficients  
$\tilde\ka_{e+}$, $\tilde\ka_{e-}$
with a sensitivity of order $10^{-14\pm 1}$
and combinations of $\tilde\ka_{o+}$, $\tilde\ka_{o-}$
with a sensitivity of order $10^{-10\pm 1}$.

The Hils-Hall experiment seeks changes 
in the beat frequency
as the velocity of the laboratory varies 
with the Earth's rotation.
The analysis assumes that experiments of the Michelson-Morley type
exclude observable sensitivity to the orientation $\th$ of the cavity.
In the context of Eq.\ \rf{ffs},
this corresponds to assuming negligible $B$ and $C$ terms.
The Fourier amplitude at the sidereal frequency is analyzed,
obtaining a bound of $2 \times 10^{-13}$
at the 90\% confidence level.
With the configuration leading to Eq.\ \rf{dnu11},
this bound constrains the combination $\sqrt{A_1^2+A_2^2}$.

Since at present many combinations of $B$ and $C$ remain unconstrained,
the assumption of negligible $B$, $C$ terms is undesirable
in the analysis of Kennedy-Thorndike experiments.
If this assumption is relaxed,
the Fourier amplitude at the sidereal frequency
contains contributions from
$A_1$, $A_2$, $B_1$, $B_2$, $C_1$, $C_2$.
It should therefore be possible to measure
combinations of the coefficients 
$\tilde\ka_{e+}$, $\tilde\ka_{e-}$
at the level of about $10^{-13}$
and combinations of the coefficients 
$\tilde\ka_{o+}$, $\tilde\ka_{o-}$
at the level of about $10^{-9}$.
A complete analysis could also 
analyze the second Fourier amplitude,
which would provide a measurement of a combination
of $A_3$, $A_4$, $B_3$, $B_4$, $C_3$, $C_4$.

The analysis of Braxmaier {\it et al.} focuses 
on a variation in $\de\nu/\nu$ with the orbital motion of the Earth.
The assumption of negligible $B$, $C$ terms is again made.
The data are averaged daily,
and a bound on the fractional frequency shift 
of $4.8\pm 5.3\times 10^{-12}$
is obtained.
In the context of Eq.\ \rf{dnu11},
the analysis restricts the sensitivity to $A_0$.
Using Eq.\ \rf{A} in Appendix \ref{signalcoeff},
it then follows that the reported constraint 
corresponds to a bound on a combination
of $\tilde\ka_{o+}$, $\tilde\ka_{o-}$
at the level of about $10^{-8}$.
Sensitivity to $C_0$ could also be obtained
if $B$ and $C$ terms were included.
Note that the polarization chosen in deriving Eq.\ \rf{dnu11}
implies the coefficient $B_0$ is independent of $\be_\oplus$
and hence cannot be extracted.

The above discussion shows that many interesting prospects remain 
for measurements of the coefficients $\tilde\ka$ 
using optical cavities.
Note that the published experimental analyses 
to date are each sensitive to different combinations 
of coefficients for Lorentz violation.
A systematic analysis could in principle provide sensitivity 
to all the coefficients 
$\tilde\ka_{e+}$ and $\tilde\ka_{e-}$
at the level of about $10^{-13}$ or better
and suppressed sensitivity to the coefficients 
$\tilde\ka_{o+}$ and $\tilde\ka_{o-}$
at the level of about $10^{-9}$ or better.
Note also that an analysis along the above lines 
could readily be applied
to space-based tests involving optical cavities,
including ones on the ISS
or in dedicated missions 
such as the proposed OPTIS experiment
\cite{optis}.
Some potential advantages of space-based tests are discussed below
in the context of experiments using microwave cavities.

\subsection{Microwave cavities}
\label{miccav}

Microwave-cavity oscillators are among the most stable clocks,
and as such they offer interesting prospects for Lorentz tests.
In particular,
there has recently been renewed interest
in superconducting cavity-stabilized oscillators
as clocks for use on the ISS \cite{sumo}.
Superconducting cavities made of niobium
have achieved $Q$ factors of $10^{11}$ or better,
and frequency stabilities of $3 \times 10^{-16}$
have been demonstrated.
In this section, 
we focus on perturbations of microwave-cavity resonant frequencies
arising from the coefficients $\kfi$ for Lorentz violation.

\subsubsection{Theory}

Equation \rf{dnu1} can be applied
to obtain the fractional resonant-frequency shift $\de\nu/\nu$
for a superconducting microwave cavity of any geometry.
The highest $Q$ factors have been demonstrated in
cylindrical cavities with circular cross section,
so we focus on this case.
For simplicity,
we suppose the cavity contains a medium of
relative transverse permittivity $\ep$,
relative axial permittivity $\ep '$,
and relative permeability $\mu = 1$.
The vacuum case is recovered as the limit $\ep = \ep ' = 1$.

The invariance of the cylindrical geometry 
under a parity transformation 
suggests the matrix ${\cal M}_{DB}$ vanishes,
since the coefficients $(\ka_{DB})^{jk}$ are parity odd. 
Also, the cavity is invariant under rotations about the symmetry axis,
so in a cavity frame with 3 axis along the symmetry axis
we expect the rotational symmetry to imply diagonal 
matrices $({\cal M}_{DE})_{\rm cav}$ and $({\cal M}_{HB})_{\rm cav}$
with equal $\{11\}$ and $\{22\}$ components.
Indeed, 
for both TE$_{mnp}$
and TM$_{mnp}$ modes,
we obtain 
\bea
\fr{\de\nu}\nu&=&
({\cal M}_{DE})_{\rm cav}^{11}[(\ka_{DE})_{\rm cav}^{11}
+(\ka_{DE})_{\rm cav}^{22}]
\nonumber \\
&&
+({\cal M}_{DE})_{\rm cav}^{33}(\ka_{DE})_{\rm cav}^{33} 
\nonumber \\
&&
+({\cal M}_{HB})_{\rm cav}^{11}[(\ka_{HB})_{\rm cav}^{11}
+(\ka_{HB})_{\rm cav}^{22}] 
\nonumber \\
&&
+({\cal M}_{HB})_{\rm cav}^{33}(\ka_{HB})_{\rm cav}^{33} 
\label{dnumicro1}
\eea
in the cavity frame.

For the TM$_{mnp}$ modes,
some calculation reveals that 
the nonzero elements of the ${\cal M}$ matrices are:
\bea
({\cal M}_{DE})_{\rm cav}^{11}&=&
({\cal M}_{DE})_{\rm cav}^{22}
=-\frac14\fr{\ep'(\pi pR)^2}
{\ep\ep'(\pi pR)^2+\ep^2(x_{mn}d)^2} , 
\nonumber \\
({\cal M}_{DE})_{\rm cav}^{33}&=&
-\half \fr{\ep(x_{mn}d)^2}{\ep'^2(\pi pR)^2+\ep\ep'(x_{mn}d)^2} ,
\nonumber \\
({\cal M}_{HB})_{\rm cav}^{11}&=&
({\cal M}_{HB})_{\rm cav}^{22}=\frac14 ,
\label{Mcav}
\eea
where $R$ and $d$ are the radius and length of
the cavity,
and where $x_{mn}$ is the $n$th zero of
the $m$th-order Bessel function $J_m(x)$.
The corresponding results for the TE$_{mnp}$ modes are
\bea
({\cal M}_{DE})_{\rm cav}^{11}&=&
({\cal M}_{DE})_{\rm cav}^{22}=-\fr1{4\ep} ,
\nonumber \\
({\cal M}_{HB})_{\rm cav}^{11}&=&
({\cal M}_{HB})_{\rm cav}^{22}
 =\frac14\fr{(\pi pR)^2}{(\pi pR)^2+(x'_{mn}d)^2} , 
 \nonumber \\
({\cal M}_{HB})_{\rm cav}^{33}&=&
\half \fr{(x'_{mn}d)^2}{(\pi pR)^2+(x'_{mn}d)^2} ,
\label{Ecav}
\eea
where $x'_{mn}$ is the $n$th zero of the
derivative of $J_m(x)$.
Note that taking the optical-cavity limit
$p\rightarrow\infty$
of any TM$_{mnp}$ or TE$_{mnp}$ mode 
yields a result identical to that obtained by 
averaging over all optical-cavity polarizations 
in Eq.\ \rf{opticalMs}.

For practical applications,
it is useful to generalize 
Eq.\ \rf{dnumicro1}
to the case where the cavity is arbitrarily oriented
in one of the standard laboratory frames
introduced in Appendix \ref{frames}.
In the laboratory frame,
denote the components of a unit vector 
parallel to the symmetry axis of the cavity
by $\hat N^j$.
The fractional frequency shift is then found to be
\bea
\fr{\de\nu}\nu&=&
({\cal M}_{DE})_{\rm cav}^{11}(\ka_{DE})_{\rm lab}^{jj}
+({\cal M}_{HB})_{\rm cav}^{11}(\ka_{HB})_{\rm lab}^{jj} 
\nonumber \\
&&
+[({\cal M}_{DE})_{\rm cav}^{33}-({\cal M}_{DE})_{\rm cav}^{11}]
\hat N^j\hat N^k(\ka_{DE})_{\rm lab}^{jk} 
\nonumber \\
&&
+[({\cal M}_{HB})_{\rm cav}^{33}-({\cal M}_{HB})_{\rm cav}^{11}]
\hat N^j\hat N^k(\ka_{HB})_{\rm lab}^{jk} .
\label{dnumicro2}
\eea
This implies the relationships
\bea
({\cal M}_{DE})_{\rm lab}^{jk}&=&
[({\cal M}_{DE})_{\rm cav}^{33}-({\cal M}_{DE})_{\rm cav}^{11}]
\hat N^j\hat N^k 
\nonumber \\
&&
+({\cal M}_{DE})_{\rm cav}^{11}\de^{jk} ,
\nonumber \\
({\cal M}_{HB})_{\rm lab}^{jk}&=&
[({\cal M}_{HB})_{\rm cav}^{33}-({\cal M}_{HB})_{\rm cav}^{11}]
\hat N^j\hat N^k 
\nonumber \\
&&
+({\cal M}_{HB})_{\rm cav}^{11}\de^{jk} ,
\nonumber \\
({\cal M}_{HB})_{\rm lab}^{jk}&=&0 .
\label{Mlab}
\eea

Using Eq.\ \rf{dnumicro2} and 
the transformation \rf{trans},
we can write the fractional frequency shift for a general mode
in terms of coefficients for Lorentz violation
in the Sun-centered celestial equatorial frame.
To order $\be$, 
we find
\bea
\fr{\de\nu}{\nu}
&=&
-\frac14\hat N^j\hat N^kR^{jJ}R^{kK}(\tilde\ka_{e'})^{JK}
-\frac14({\cal M}_2+3{\cal M}_3)\tilde\ka_{\rm tr}
\nonumber \\
&&
\hskip -5 pt
-\half ({\cal M}_3\de^{jk}/{\cal M}_2 +\hat N^j\hat N^k )
R^{jJ}R^{kK}\ep^{JPQ}\be^Q (\tilde\ka_{o'})^{KP}.
\nonumber\\
\label{gendnu}
\eea
In this equation,
we define the quantities
\bea
{\cal M}_1&\equiv& -4[
({\cal M}_{DE})_{\rm cav}^{33}
-({\cal M}_{DE})_{\rm cav}^{11}
\nonumber \\ && \quad
+({\cal M}_{HB})_{\rm cav}^{33}
-({\cal M}_{HB})_{\rm cav}^{11}] ,
\nonumber \\
{\cal M}_2&\equiv& -4[
({\cal M}_{DE})_{\rm cav}^{33}
-({\cal M}_{DE})_{\rm cav}^{11}
\nonumber \\ && \quad
-({\cal M}_{HB})_{\rm cav}^{33}
+({\cal M}_{HB})_{\rm cav}^{11}] ,
\nonumber \\
{\cal M}_3&\equiv& -4[
({\cal M}_{DE})_{\rm cav}^{11}
-({\cal M}_{HB})_{\rm cav}^{11}],
\label{calms}
\eea
which depend on the cavity mode
and control the linear combinations
\bea
(\tilde\ka_{e'})^{JK}&=&
{\cal M}_1(\tilde\ka_{e+})^{JK}
+{\cal M}_2(\tilde\ka_{e-})^{JK} ,
\nonumber \\
(\tilde\ka_{o'})^{JK}&=&
{\cal M}_1(\tilde\ka_{o-})^{JK}
+{\cal M}_2(\tilde\ka_{o+})^{JK}
\label{convenient}
\eea
of coefficients for Lorentz violation.
These equations reveal that
the sensitivity 
of experiments with microwave cavities
to Lorentz violation 
varies with the mode and with the
permittivity of the medium in the cavity.
As before, 
to this order $\tilde\ka_{\rm tr}$
contributes only to an unobservable constant frequency shift.

As an illustration,
consider a cavity void of matter
and operated on the fundamental TM$_{010}$ mode,
as planned for some space- and ground-based experiments.
For this case,
we find ${\cal M}_1 = 3$, ${\cal M}_2 = {\cal M}_3 = 1$,
and the fractional frequency shift \rf{gendnu} becomes
\bea
\fr{\de\nu}{\nu} \Bigr\vert_{{TM}_{010}}
&=&
-\frac 14 \hat N^j \hat N^k
[2 (\ka_{DE})_{\rm lab}^{jk} + (\ka_{HB})_{\rm lab}^{jk}
\nonumber\\
&&
\qquad\qquad\qquad
- \de^{jk} (\ka_{HB})_{\rm lab}^{ll}]
\nonumber\\
&=&
-\frac14\hat N^j\hat N^kR^{jJ}R^{kK}
[3(\tilde\ka_{e+})^{JK}+(\tilde\ka_{e-})^{JK}]
\nonumber \\
&&
-\half (\de^{jk}+\hat N^j\hat N^k ) R^{jJ}R^{kK}\ep^{JPQ}\be^Q 
\nonumber \\
&&\qquad \times
[3(\tilde\ka_{o-})^{KP}+(\tilde\ka_{o+})^{KP}]
-\tilde\ka_{\rm tr}.
\label{dnuTM010}
\eea
The observable shift depends on the traceless symmetric matrix
combination 
$3(\tilde\ka_{e+})^{JK}+(\tilde\ka_{e-})^{JK}$
and the traceless matrix combination 
$3(\tilde\ka_{o-})^{JK}+(\tilde\ka_{o+})^{JK}$.
The first of these contains five 
linearly independent combinations of the 11 parity-even coefficients
for Lorentz violation, 
while the second contains all eight parity-odd coefficients.
Note that certain harmonics may be sensitive to smaller
subset of these 13 quantities.
For example, for a fixed Earth-based cavity,
if $\hat\om$ represents the rotational axis of
the Earth's revolution,
then the sidereal harmonics are insensitive to the component of
$3(\tilde\ka_{e+})^{JK}+(\tilde\ka_{e-})^{JK}$
proportional to $\hat\om^J\hat\om^K$, reducing
the number of combinations to 12.

If an Earth-based experiment is performed over a period of time 
$\De T_{\rm exp}$
short compared to the orbital period of the Earth,
then the velocity $\be$ is roughly constant
and the experiment is sensitive
primarily to the four linear combinations corresponding to the
vector amplitudes of the two harmonics.
To acquire sensitivity to other combinations,
the Earth-based experiment could either
be repeated several times during the year,
or the cavity could be rotated in the laboratory.
In contrast, 
for a satellite-based experiment, 
perturbations cause the orbital plane 
and hence the analogue of $\hat \om$ to precess with time.
Also, the smaller orbital period implies different harmonics 
and access to more coefficients for the same $\De T_{\rm exp}$.
As a result,
if the experiment is performed two or more times
with significantly different orbital planes,
all 13 combinations of coefficients can be
accessed through the orbital-frequency harmonics.

Equation \rf{dnumicro2} can be adapted to either
a space-based or Earth-based experiment
and, if necessary,
to the case of a rotating cavity.
In the remainder of this section,
we offer some remarks about possible experiments 
with microwave cavities on the ISS
and on the Earth.

\subsubsection{Space-based experiment}
\label{spaceexpt}

The construction of the ISS offers the possibility 
of performing Lorentz tests in low Earth orbit.
Of particular relevance in the present context is
the SUMO experiment 
\cite{sumo},
which plans to use superconducting microwave-cavity oscillators
as clocks on upcoming ISS flight missions.

The ISS operates in several different flight modes,
which correspond to different laboratory configurations
in the Sun-centered celestial equatorial frame.
Each flight mode therefore involves different transformations 
\rf{trans},
which could lead to different 
sensitivities to the Lorentz-violating coefficients.
If,
for example, 
the ISS orientation were fixed in Sun-centered frame,
corresponding to no rotation during an orbit,
then the signal would involve only boost-dependent terms
with a 92-minute period.
For definiteness and simplicity,
we focus here on a flight mode
with the ISS $z$-axis aligned along its orbital velocity
with respect to the Earth.
This corresponds to the standard laboratory frame 
introduced in Appendix \ref{frames}.

For a microwave cavity with fixed orientation 
$\hat N$ in this ISS laboratory frame,
several harmonics could be studied.
The fractional frequency shift ${\de\nu}/\nu$
varies with the Earth's orbital frequency, 
the ISS orbital frequency $\om_s$,
and the ISS orbital-precession frequency.
The most interesting of these 
is likely to be the highest frequency, $\om_s$.

In practice,
the fractional frequency shift may be  
measured relative to another oscillator clock
via the beat frequency of the combined signal.
The reference clock 
could be a different physical system,
such as a hydrogen maser or atomic clock,
which could conveniently be operated
on a transition known to be insensitive
to Lorentz violation
\cite{bklr}.
A comparison of two microwave cavities could also be used.
For example,
SUMO may involve a pair of cavities 
oriented at right angles to each other.
The observed signal would then depend strongly 
on the orientation of the pair in the ISS frame.
Thus,
at leading order in $\be$,
a cavity oriented with $\hat N$ 
perpendicular to the orbital plane 
is insensitive to the parity-even coefficients
for Lorentz violation,
and only $\be$-suppressed parity-odd terms appear in the
frequency shift.
In contrast,
a cavity positioned 
with $\hat N$ in the orbital plane maximizes the
sensitivity in the second harmonics of $\om_s$, 
while one with $\hat N$ at $45^\circ$ from the orbital plane 
maximizes the first harmonics.
These results can be obtained directly from Eq.\ \rf{dnumicro2}.

For a pair of identical cavities,
the variation in the beat frequency takes the general form
\bea
\fr{\nu_{beat}}{\nu} &\equiv&  \fr {\de\nu_1} {\nu} - \fr{\de\nu_2}{\nu}
\nonumber \\ 
&=& {\cal A}_s\sin\om_sT_s +{\cal A}_c\cos\om_sT_s 
\nonumber \\
&&
+{\cal B}_s\sin2\om_sT_s+{\cal B}_c\cos2\om_sT_s + {\cal C} ,
\label{dnu7}
\eea
where
${\cal A}_s$, ${\cal A}_c$, ${\cal B}_s$, and ${\cal B}_c$
are four linear combinations of the coefficients $\kfi$
for Lorentz violation.
These combinations depend on the orientations 
$\hat N_1$, $\hat N_2$ of the cavity pair
and on the orientation of the orbital plane 
with respect to the Sun-centered celestial equatorial frame.
The precession of the ISS orbit slowly changes the four combinations,
allowing access to more coefficients.
Typically,
the combinations are rather cumbersome.
Appendix \ref{spacecoeff} contains their explicit form 
for a maximal-sensitivity case,
for which $\hat N_1 = (0,0,1)$ and $\hat N_2 = (1,1,0)/\sqrt{2}$.
The expressions involve the linear combinations
\rf{convenient}, 
which hold for an arbitrary mode and arbitrary permittivities
$\ep$, $\ep '$.

\subsubsection{Earth-based experiment}

For an Earth-based experiment with a cavity pair fixed
in the laboratory,
the dominant frequency is 
the Earth's sidereal frequency $\om_\oplus$.
The equivalent of the ISS orbital plane in this case
is the plane in which the laboratory moves,
which parallels the equatorial plane 
at the latitude of the laboratory.
As before,
the configuration of maximum sensitivity 
has one cavity in this plane 
and the other at $45^\circ$ to it.
However, 
for laboratories located in middle latitudes,
it suffices to orient one cavity horizontally 
in the east-west direction
and the other either vertically 
or horizontally in the north-south direction.
The east-west cavity is then maximally sensitive to the
second harmonics, 
while the north-south cavity is near maximal sensitivity 
to the first harmonics.
The latter are proportional to $\cos2\ch$,
so for colatitudes in the range $30^\circ<\ch<60^\circ$
there is at most a 14\% reduction in sensitivity.

For definiteness,
we consider the configuration with the second cavity
oriented vertically in the laboratory.
The laboratory-frame orientation vectors are then 
$\hat N_1=(0,1,0)$ and $\hat N_2=(0,0,1)$.
Paralleling the discussion leading to Eq.\ \rf{dnu7},
we write the fractional beat frequency 
due to Lorentz violation as
\bea
\fr{\nu_{beat}}\nu &=& 
{\cal A}_{\oplus s}\sin\om_\oplus T_\oplus
+{\cal A}_{\oplus c}\cos\om_\oplus T_\oplus 
\nonumber \\
&&
+{\cal B}_{\oplus s}\sin2\om_\oplus T_\oplus
+{\cal B}_{\oplus c}\cos2\om_\oplus T_\oplus + {\cal C}_\oplus .
\label{dnu8}
\eea
At first order in $\be$, 
we find 
\bea
{\cal A}_{\oplus s}&=&\frac14\sin2\ch(\tilde\ka_{e'})^{YZ}
\nonumber\\
&&
-\frac14\be_\oplus\sin2\ch
\bigl[\sin\Om_\oplus T
\bigl((\tilde\ka_{o'})^{YY}-(\tilde\ka_{o'})^{ZZ}\bigr)
\nonumber\\
&&
-\sin\et\cos\Om_\oplus T(\tilde\ka_{o'})^{ZX}
+\cos\et\cos\Om_\oplus T
(\tilde\ka_{o'})^{YX}\bigr]
\nonumber\\
&&
-\half\be_L\bigl(\sin^2\ch(\tilde\ka_{o'})^{YZ}
-\cos^2\ch(\tilde\ka_{o'})^{ZY}\bigr) ,
\nonumber\\
{\cal A}_{\oplus c}&=&\frac14\sin2\ch(\tilde\ka_{e'})^{XZ}
-\frac14\be_\oplus\sin2\ch
\nonumber\\
&&
\times\bigl[\sin\Om_\oplus T(\tilde\ka_{o'})^{XY}
+\sin\et\cos\Om_\oplus T(\tilde\ka_{o'})^{ZY}
\nonumber\\
&&
+\cos\et\cos\Om_\oplus T
\bigl((\tilde\ka_{o'})^{XX}-(\tilde\ka_{o'})^{ZZ}\bigr)\bigr]
\nonumber\\
&&
-\half\be_L\bigl(\sin^2\ch(\tilde\ka_{o'})^{XZ}
-\cos^2\ch(\tilde\ka_{o'})^{ZX}\bigr) ,
\nonumber\\
{\cal B}_{\oplus s}&=&\frac14\bigl(1+\sin^2\ch\bigr)(\tilde\ka_{e'})^{XY}
+\frac14\be_\oplus\bigl(1+\sin^2\ch\bigr)
\nonumber\\
&&
\times\bigl[\sin\Om_\oplus T(\tilde\ka_{o'})^{XZ}
+\cos\et\cos\Om_\oplus T(\tilde\ka_{o'})^{YZ}
\nonumber\\
&&
+\sin\et\cos\Om_\oplus T
\bigl((\tilde\ka_{o'})^{XX}-(\tilde\ka_{o'})^{YY}\bigr)\bigr]
\nonumber\\
&&
+\frac18\be_L\sin2\ch\bigl((\tilde\ka_{o'})^{XY}
+(\tilde\ka_{o'})^{YX}\bigr) ,
\nonumber\\
{\cal B}_{\oplus c}&=&\frac18\bigl(1+\sin^2\ch\bigr)
\bigl((\tilde\ka_{e'})^{XX}
-(\tilde\ka_{e'})^{YY}\bigr)
\nonumber\\ 
&&
-\frac14\be_\oplus\bigl(1+\sin^2\ch\bigr)
\nonumber\\
&&
\times\bigl[\sin\Om_\oplus T(\tilde\ka_{o'})^{YZ}
-\cos\et\cos\Om_\oplus T(\tilde\ka_{o'})^{XZ}
\nonumber\\
&&
+\sin\et\cos\Om_\oplus T
\bigl((\tilde\ka_{o'})^{XY}+(\tilde\ka_{o'})^{YX}\bigr)\bigr]
\nonumber\\
&&
+\frac18\be_L\sin2\ch\bigl((\tilde\ka_{o'})^{XX}
-(\tilde\ka_{o'})^{YY}\bigr) ,
\nonumber\\
{\cal C}_\oplus &=&\frac38\cos^2\ch(\tilde\ka_{e'})^{ZZ}
\nonumber\\ 
&&
-\frac14\be_\oplus\cos^2\ch
\bigl[\sin\Om_\oplus T
\bigl((\tilde\ka_{o'})^{YZ}+2(\tilde\ka_{o'})^{ZY}\bigr)
\nonumber\\
&&
-\sin\et\cos\Om_\oplus T
\bigl((\tilde\ka_{o'})^{XY}-(\tilde\ka_{o'})^{YX}\bigr)
\nonumber\\
&&
+\cos\et\cos\Om_\oplus T
\bigl((\tilde\ka_{o'})^{XZ}+2(\tilde\ka_{o'})^{ZX}\bigr)\bigr]
\nonumber\\
&&
-\frac38\be_L\sin2\ch(\tilde\ka_{o'})^{ZZ} ,
\eea
where the convenient combinations \rf{convenient}
have been adopted.
As before,
these equations are valid for any specific mode 
and for arbitrary permittivities $\ep$, $\ep '$. 

A Lorentz-violating signal 
would thus manifest itself as a sidereal variation in the
fractional beat frequency according to Eq.\ \rf{dnu8}.
At zeroth order in $\be$,
the corresponding amplitude associated
with this variation is constant 
and determined by the four parity-even coefficients 
$(\tilde\ka_{e'})^{XZ}$,
$(\tilde\ka_{e'})^{YZ}$,
$(\tilde\ka_{e'})^{XY}$,
and $(\tilde\ka_{e'})^{XX}-(\tilde\ka_{e'})^{YY}$.
These linearly independent combinations of $\kfi$
remain unmeasured to date.
When the first-order terms in $\be$ are included, 
the amplitudes also contain harmonics at the Earth's orbital frequency
$\Om_\oplus$.
The resulting variations depend on the eight parity-odd coefficients
$\tilde\ka_{o'}$.
The three of these represented by $\tilde\ka_{o+}$ have yet to
be measured.

The above experiment provides access to the specified 
parity-even coefficients at the level of the cavity stability,
which for available microwave cavities
could be at the order of $10^{-13}$ or better.
The $\be$ suppression reduces the sensitivity to
the parity-odd coefficients to about 
$10^{-9}$ or better.
The experiments can be performed on any cavity mode
and for cavities with or without matter.
For example,
as can be seen from the explicit expressions
\rf{convenient},
a pair of sapphire-filled cavities
($\ep\simeq 9.5$, $\ep'\simeq 11.5$)
operated on a whispering-gallery mode
\cite{wd}
offers sensitivity to linear combinations
of coefficients for Lorentz violation
that differ from those of a pair of vacuum cavities 
operated on the fundamental TM$_{010}$ mode.
In fact,
for any specified combination of cavities with known fields,
the matrices $({\cal M}_{DE})_{\rm cav}$, $({\cal M}_{HB})_{\rm cav}$
in Eq.\ \rf{dnumicro1}
can be determined 
and hence the fractional beat frequency can be obtained as above.
Note also that other coefficients could be accessed
by placing the cavity pair on a rotating turntable,
which would also allow a dataset to be obtained 
in weeks or days rather than months.

\subsection{Cavity Deformation}

In this remaining subsection,
we offer a few remarks concerning the possibility 
that Lorentz violation might alter atomic binding forces 
and hence the structural properties of matter.
In particular,
one resulting effect relevant to cavity experiments
might be a deformation of the cavity,
which could change the resonant frequency.
The issue is whether possible deformations arising from 
Lorentz-violating effects in the standard-model extension
could cancel the predicted effects.
The differences between the transformation properties and
nature of the various coefficients for Lorentz violation make
it unlikely that coefficients other than
$\kfi$ could cause complete cancellation of the signals
discussed above,
so a more interesting question is whether the coefficients
$\kfi$ could alter the dimensions of a cavity
so as to offset completely the predicted signals.

Any leading-order modifications 
to atomic and molecular binding forces 
arising from nonzero $\kfi$ 
are expected to come from modifications to the Coulomb potential.
The form of the Gauss law \rf{max2}
in the presence of Lorentz violation due to $\kfi$
implies that the modified Coulomb potential 
for a point charge $q$ is
\beq
\Ph(\vec x) = \fr{q}{4\pi|\vec x|}\left(1
  +\fr{\vec x\cdot\ka_{DE}\cdot\vec x}{2\vec x^2} \right) .
\eeq
The leading-order effects on the physical dimensions 
of the cavity are therefore expected to depend only 
on the matrix $\ka_{DE}$.
We could account phenomenologically for such effects
by adding a term to Eq.\ \rf{dOlab} of the form
${\cal M}_{\rm matter}^{jk}(\ka_{DE})_{\rm lab}^{jk}$,
where the constant matrix ${\cal M}_{\rm matter}^{jk}$ 
is determined by the properties of the material from which 
the cavity is made.
For example, in a simple ionic lattice model,
${\cal M}_{\rm matter}^{jk}$ depends on the charge of the ions, 
the lattice configuration,
and the orientation of the cavity with respect to the lattice.

For the optical and microwave cavities considered here, 
this extra term cannot completely cancel 
the predicted fractional frequency shifts.
Although a partial cancellation might be possible in principle,
it requires that the matrix ${\cal M}_{\rm matter}^{jk}$ 
takes a special form that is improbable in light of the complexity 
of the binding forces of solids.

\section{SUMMARY}
\label{summary}

In this work, 
we studied the Lorentz-violating electrodynamics 
derived from the renormalizable sector of the full 
Lorentz-violating standard-model extension.
Some basic material is presented in
Sec.\ \ref{theory},
followed by a useful analogy and some definitions 
in Sec.\ \ref{defs}.
Sec.\ \ref{others} 
discusses the connection to some test models.

The bulk of the paper is devoted to tests of 
the Lorentz-violating electrodynamics
and methods to measure the 19 independent coefficients $\kfi$
for Lorentz violation.
We first consider astrophysical tests
based on the prediction that the vacuum is birefringent.
Theoretical issues pertaining to vacuum birefringence
are discussed in Sec.\ \ref{biretheory}.
One potentially observable effect 
is the dispersion of pulses over astrophysical distances.
The constraint \rf{velbound} on $\kfi$ from this effect 
is obtained in Sec.\ \ref{velocity}.
Another potentially observable effect arises in the 
comparative spectropolarimetry of cosmological sources.
The tight bound \rf{polbound} on 10 of the 19 coefficients $\kfi$
is obtained in Sec.\ \ref{polar}.

The possibility of sensitive laboratory tests 
of Lorentz invariance is examined 
in Sec.\ \ref{labtheory}.
A general framework for the analysis of both Earth-based
and space-based experiments is provided.
The analysis is applied to two types of cavity-stabilized
oscillator experiments.
In Sec.\ \ref{optcav}, 
we consider optical-cavity experiments.
High sensitivity microwave-cavity experiments are discussed
in Sec.\ \ref{miccav}.
We find that appropriate 
laboratory tests can access all 19 coefficients $\kfi$.

Table 3 summarizes the existing constraints.
The 19 coefficients $\kfi$ are represented by the 
matrices $\tilde\ka_{e+}$, $\tilde\ka_{e-}$,
$\tilde\ka_{o+}$, $\tilde\ka_{o-}$, $\tilde\ka_{tr}$
defined in Eq.\ \rf{kappas2}.
The number of independent components in each matrix
is shown in the second column.
The order of magnitude of the astrophysical bounds
is shown in the third and fourth column.
These bounds tightly constrain the 10 coefficients $\kfi$ contained
in $\tilde\ka_{e+}$ and $\tilde\ka_{o-}$.
However,
as indicated in the table by the notation n/a,
the remaining coefficients are unobservable 
in astrophysical tests.
In contrast, 
laboratory experiments with optical and microwave cavities
can in principle access all the coefficients.
As discussed in Sec.\ \ref{optcav},
several recent experiments with optical cavities
\cite{bh,hh,brax}
offer sensitivity to a few of the coefficients 
at levels lying between about $10^{-8}$ and $10^{-15}$,
but no definitive analysis has been performed.
The matrices for which a few components are probably constrained 
in this way are indicated by the symbol $\star$ in the table.
To date,
no measurements of Lorentz violation
using microwave cavities have been reported. 

\begin{center}
\begin{tabular}{|l|c||c|c||c|c|}
\hline
\hline
\multicolumn{2}{|c||}{ }
& \multicolumn{2}{c||}{ Astrophysical Tests}
& \multicolumn{2}{c|}{ Laboratory Tests}
\\
\hline
\hline
Coeff.\ & No.\ & Velocity & Polarization & Optical & Microwave \\
\hline
\quad
$\tilde\ka_{e+}$& 5 & -16 & -32  & $\star$ & - \\
\quad
$\tilde\ka_{e-}$& 5 & n/a & n/a & $\star$ & - \\
\quad
$\tilde\ka_{o+}$& 3 & n/a & n/a & $\star$ & - \\
\quad
$\tilde\ka_{o-}$& 5 & -16 & -32 & $\star$ & - \\
\quad
$\tilde\ka_{\rm tr}$& 1 & n/a & n/a & - & - \\
\hline
\hline
\end{tabular}
\end{center}
\begin{center}
Table 3. Existing constraints.
\label{const}
\end{center}

In conclusion,
astrophysical observations 
place bounds on Lorentz violation in electrodynamics
that are competitive with ones in the fermion sectors
obtained by other means.
Laboratory experiments are needed to complement these measurements 
by spanning the allowed parameter space in the photon sector,
and the technology presently exists to perform them.
These experiments offer a promising avenue 
to search for new physics lying beyond the standard model.

\acknowledgments
We thank 
Steve Biller, John Dick, John Lipa, 
Joel Nissen, Subir Sarkar, and Ron Walsworth
for discussions.  
This work was supported in part by the 
National Aeronautics and Space Administration
under grant number NAG8-1770 and by the 
United States Department of Energy
under grant number DE-FG02-91ER40661.

\appendix

\section{Birefringence Vectors}
\label{birefvec}

For a distant source viewed from the Earth at declination
$d$ and right ascension $r$, 
the direction of propagation towards the Earth
can be written as
$\pht^{\,\mu}=(1;-\cosd\cosr,-\cosd\sinr,-\sind)$.
The matrix 
\beq
R^{jK}=\left(
\begin{array}{ccc}
\sind\cosr & \sind\sinr & -\cosd \\
\sinr & -\cosr & 0 \\
-\cosd\cosr & -\cosd\sinr & -\sind
\end{array}
\right)
\label{rot}
\eeq
implements the rotation between the primed frame
and the standard Sun-centered frames.
With this definition, 
the primed-frame basis vector $\hat e'_3$ 
points from the source towards the Earth.
The vectors $\hat e'_1$ and $\hat e'_2$ point
south and west,
respectively.

Writing $\si\sin\xi$ and $\si\cos\xi$ in terms
of coefficients in the Sun-centered celestial equatorial frame
gives 
\bea
\si\sin\xi &=& \half(R^{1J}R^{2K}+R^{2J}R^{1K})
\kf^{J\mu K\nu} \pht_\mu \pht_\nu,
\nonumber \\
\si\cos\xi &=& \half(R^{1J}R^{1K}-R^{2J}R^{2K})
\kf^{J\mu K\nu} \pht_\mu \pht_\nu .
\label{xi1}
\eea
Note that $\xi$ is not a rotational scalar,
unlike $\rh$ and $\si$.

The rotation \rf{rot} can be substituted in this result
to yield $\si\sin\xi$ and $\si\cos\xi$ in terms of $\kfi$ 
in Sun-centered celestial equatorial coordinates.
The relevant combinations of the $\kfi$ are
the 10 coefficients $k^a$ given in Eq.\ \rf{ka}.
It is convenient to express $\si\sin\xi$ and $\si\cos\xi$ 
as the scalar product of $k^a$ 
with two 10-dimensional vectors.
Defining
\bea
\vs_s^a =
\left(
\begin{array}{c}
\cosds + \cosrs - \sinds \sinrs \\
\sinds \cosrs - \cosds - \sinrs \\
-2 \sind \sinr \cosr \\
-\sind \sinr \cosr \\
\sind ( \sinrs - \cosrs ) \\
-\cosd \sinr \\
\cosd \cosr \\
-\sind \cosd \cosr \\
-\cosds \sinr \cosr \\
-\sind \cosd \sinr
\end{array}
\right) \ , \\
\vs_c^a =
\left(
\begin{array}{c}
-2\sind \sinr \cosr \\
-2\sind \sinr \cosr \\
\half(1 + \sinds )( \sinrs - \cosrs ) \\
\half(\sind + \sinrs - \sinds \cosrs ) \\
(1 + \sinds ) \sinr \cosr \\
-\sind \cosd \cosr \\
-\sind \cosd \sinr \\
\cosd \sinr \\
\sind ( \sinrs - \cosrs ) \\
-\cosd \cosr
\end{array}
\right) \ ,
\label{vectors}
\eea
we find 
\bea
\si\sin\xi &=& \vs_s^a k^a,
\nonumber\\
\si\cos\xi &=& \vs_c^a k^a.
\eea

\section{Polarization Review}
\label{polbackground}

Conventionally, 
polarization is defined by the behavior 
of the electric field vector \cite{bw}.
The polarization of a general plane wave can be described
by an ellipse residing in the plane perpendicular
to the direction of propagation.
In terms of the primed-frame variables introduced
in Sec.\ \ref{biretheory}
and to leading order in $\kfi$, 
this plane is spanned by the basis vectors
$\hat e'_1$ and $\hat e'_2$. 
The orientation and shape of the ellipse can be described 
by two angles, $\ps$ and $\ch$.
The angle $\ps$ determines the orientation of the ellipse
and is defined as the angle between
the major axis of the ellipse and $\hat e'_1$.
The angle $\ch$ describes the shape of the ellipse 
and the helicity of the wave,
and it is defined by $\ch = \pm\arctan r$,
where $r$ is the ratio of the minor to major axes of the ellipse.

In polarimetry, 
the ellipse is commonly parametrized using Stokes parameters.
We define a Stokes vector by
\bea
(s^0,\vec s)&\equiv&
\bigl( |E'_1|^2+|E'_2|^2,~ |E'_1|^2-|E'_2|^2,~
\nonumber\\
&&\qquad\qquad 2\Re{{E'_1}^*E'_2},~ 2\Im{{E'_1}^*E'_2}\bigr)
\nonumber\\
&=&
s^0 ( 1,~ \cos2\ch\cos2\ps,~ \cos2\ch\sin2\ps,~ \sin2\ch) . 
\nonumber\\
\eea

\begin{figure}[]
\centerline{
\psfig{figure=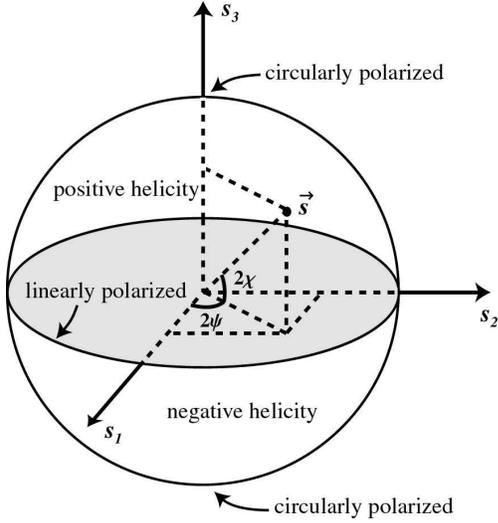,width=0.8\hsize}}
\caption{The Poincar\'e sphere.}
\label{sphere1}
\end{figure}

In the context of the discussion in Sec.\ \ref{polar},
the losslessness of the vacuum implies that the Stokes parameter 
$s^0$ is unaffected at leading order by a relative-phase change.
We therefore normalize to $s^0=1$ throughout.
With $s^0=1$, 
each Stokes vector $\vec s$ represents
a unique point on a two-dimensional sphere of unit radius,
called the Poincar\'e sphere.
As illustrated in Fig.\ \ref{sphere1},
$2\ps$ and $2\ch$ are the angles that specify the
position of $\vec s$ on this sphere.
An arbitrary polarization is represented by
a single point on the sphere.
The points in the $s^1$-$s^2$ plane represent
all linear polarizations.
The points in the upper hemisphere all represent 
elliptical polarizations of positive helicity,
with the pole being the special case of circular polarization.
Similarly, 
the lower hemisphere represents 
polarizations of negative helicity.

In Sec.\ \ref{polar},
the quantity of interest is the polarization change
induced by the phase shift $\De\ph$ in Eq.\ \rf{deph}.
The effect of $\De\ph$ on the Stokes vector $\vec s$ 
can be visualized in terms of motion on the Poincar\'e sphere.
Consider an arbitrary orthonormal elliptical basis
$\{\hat\ve_1,\hat\ve_2\}$.
The associated Stokes vectors
$\vec s_{\hat\ve_1}$, $\vec s_{\hat\ve_2}$
determine opposite points on the Poincar\'e sphere.
Decomposing a general electric field in this basis gives
polarization components $E_n e^{-i\phi_n}$, $n=1,2$,
where $E_n$ and $\phi_n$ are real.
Examining the Stokes vector for this configuration
shows that a change $\De(\ph_1-\ph_2)$ in the relative phase
results in a right-handed rotation of the Stokes vector
by the angle $\De(\ph_1-\ph_2)$ about the axis
given by $\vec s_{\hat\ve_1}=-\vec s_{\hat\ve_2}$.

\section{Standard Frames}
\label{frames}

This appendix defines our standard frames
for Earth- and space-based laboratories. 
We provide the rotations and velocities used in transforming
to the reference Sun-centered celestial equatorial frame,
which is defined in section \ref{biretheory}.

\subsection{Earth-based laboratory}

For a laboratory fixed to the surface of the Earth
in the northern hemisphere,
we choose the standard frame to have coordinates $(t,x,y,z)$ 
such that the $x$-axis points south, the $y$-axis points east,
and the $z$-axis points vertically upwards.
With the reasonable approximation that
the orbit of the Earth is circular,
the rotation from the Sun-centered celestial equatorial frame
to the standard laboratory frame is given by
\beq
R^{jJ}=\left(
\begin{array}{ccc}
\cos\ch\cos\om_\oplus T_\oplus 
& 
\cos\ch\sin\om_\oplus T_\oplus 
& 
-\sin\ch 
\\
-\sin\om_\oplus T_\oplus 
& 
\cos\om_\oplus T_\oplus 
& 
0
\\
\sin\ch\cos\om_\oplus T_\oplus 
&
\sin\ch\sin\om_\oplus T_\oplus 
& 
\cos\ch
\end{array}
\right) .
\eeq
In this equation, 
$j=$ $x,y,z$ $= 1,2,3$ denotes an index in the laboratory frame, 
while $J=X$, $Y$, $Z$ denotes an index in the Sun-centered frame. 
The Earth's sidereal angular frequency is 
$\om_\oplus\simeq 2\pi$/(23 h 56 min.),
and $\ch$ is the colatitude of the laboratory.
The time $T_\oplus$ is measured in the Sun-centered frame
from one of the times when the $y$- and $Y$-axes coincide,
to be chosen conveniently for each experiment.
The time $T_\oplus$ therefore differs 
from the celestial equatorial time $T$ 
by a constant shift for each experiment. 

The velocity 3-vector of the laboratory in the 
Sun-centered frame is
\beq
\vec\be=\be_\oplus\left(
\begin{array}{c} \sin\Om_\oplus T \\ 
-\cos\et \cos\Om_\oplus T \\ -\sin\et\cos\Om_\oplus T
\end{array}\right)+\be_L\left(
\begin{array}{c} 
-\sin\om_\oplus T_\oplus 
\\ 
\cos\om_\oplus T_\oplus 
\\ 
0 
\end{array}\right) \ .
\eeq
Here,
$\Om_\oplus$ and $\be_\oplus$ are, respectively,
the angular frequency and speed of the Earth's orbital motion.
The quantity $\eta\simeq 23.4^\circ$ 
is the angle between the $XY$ celestial equatorial plane and the
Earth's orbital plane.
The speed $\be_L=r_\oplus\om_\oplus\sin\ch\lsim 1.5\times10^{-6}$
is that of the laboratory due to the rotation of the Earth.

The reader is warned that the standard laboratory frame
defined above may differ from a frame fixed to the apparatus
in the laboratory.
For example,
the apparatus rotates in the laboratory
in some experiments considered here.
Where confusion could occur,
we distinguish with labels the quantities
defined in the standard laboratory frame 
from those in the apparatus frame.

\subsection{Space-based laboratory}

For our standard laboratory fixed to an Earth-orbiting space platform
such as the ISS,
we choose the $z$ axis 
to be aligned with the velocity $\vec \be_s$ of the satellite
with respect to the Earth.
The $x$ axis is chosen to point towards the Earth.
The $y$ axis completes a right-handed coordinate system,
thus directed along the satellite orbital angular momentum
with respect to the Earth.

The components of the matrix $R^{jJ}$ describing the rotation
from the reference Sun-centered frame 
to this standard satellite frame are
\bea
R^{1X} &=&
-\cos\al\cos\om_s T_s+\sin\al\cos\ze\sin\om_s T_s ,
\nonumber\\
R^{1Y} &=&
-\sin\al\cos\om_s T_s-\cos\al\cos\ze\sin\om_s T_s,
\nonumber\\
R^{1Z} &=&
-\sin\ze\sin\om_s T_s,
\nonumber\\
R^{2X} &=&
\sin\al\sin\ze ,
\nonumber\\
R^{2Y} &=&
-\cos\al\sin\ze ,
\nonumber\\
R^{2Z} &=&
\cos\ze,
\nonumber\\
R^{3X} &=&
-\cos\al\sin\om_s T_s-\sin\al\cos\ze\cos\om_s T_s,
\nonumber\\
R^{3Y} &=&
-\sin\al\sin\om_s T_s+\cos\al\cos\ze\cos\om_s T_s,
\nonumber\\
R^{3Z} &=&
\sin\ze\cos\om_s T_s .
\eea
Here,
$j=1,2,3$ denotes an index in the satellite frame.
The satellite orbital angular frequency is denoted $\om_s$.
The time $T_s$ is measured in the Sun-centered frame
from a conveniently chosen time when the satellite crosses 
the equatorial plane,
so the times $T_s$ and $T$ differ by a constant for each experiment.
Also,
$\ze$ is the angle between the satellite orbital plane
and the Earth's equatorial plane.
For example,
$\ze \simeq 52^\circ$ for the ISS.
The quantity $\al$ is the 
azimuthal angle at which the orbital plane
intersects the Earth's equatorial plane.
The satellite intersects the equatorial plane twice per orbit,
and $\al$ can be regarded as the angle between 
the $X$ direction and a vector from the Earth's center 
to the point where the intersection occurs on an ascending trajectory.

The velocity of the satellite with respect to
the Sun-centered celestial equatorial frame is
\bea
\vec\beta&=&\be_\oplus\left(
\begin{array}{c} \sin\Om_\oplus T \\ 
-\cos\et \cos\Om_\oplus T \\ -\sin\et\cos\Om_\oplus T
\end{array}\right) 
\nonumber\\
&&
\hskip -7 pt
+\be_s\left(\begin{array}{c}
-\cos\al \sin\om_s T_s -\sin\al \cos\ze \cos\om_s T_s \\
-\sin\al \sin\om_s T_s +\cos\al \cos\ze \cos\om_s T_s \\
\sin\ze \cos\om_s T_s
\end{array}\right).
\eea
The quantities $\be_\oplus$, $\Om_\oplus$
are defined as before.
The quantity $\be_s$ is the speed of the satellite
with respect to the Earth.
For example,
$\be_s\simeq 3\times 10^{-5}$ for the ISS.

\section{Optical frequency shift}
\label{optical}

This appendix provides an alternative method
to obtain the fractional resonant-frequency shift 
$\de\nu/\nu$ for optical cavities,
given in Eq.\ \rf{dnuopt}.
As before,
the cavity is regarded as two reflecting parallel planar surfaces
separated by a distance $l$.
For convenience, 
we let one coincide with the
$x$-$y$ plane and place the other at $z=l$.
We approximate the light entering the cavity 
as a plane wave with phase velocity parallel 
to the $z$-axis.
After each reflection inside the cavity,
the reflected wave must have the
same frequency as the incident wave,
so we consider light of constant frequency $p^0$.
For simplicity,
we set $\ep=1$ in what follows.

At leading order in the coefficients for Lorentz violation,
Eq.\ \rf{dispersion} implies that 
the magnitude of the wave vector for each birefringent mode is 
$|\vec p_\pm| = p_\pm = [1-(\rh\pm\si)]p^0$.
Decomposing the light entering the
cavity into birefringent modes,
we write 
\bea
\vec E_0(x) &=&e^{-ip^0t}
[e^{ip_{\uparrow +}z}(\hat\ve_{\uparrow +}\cdot\vec E_0)
\hat\ve_{\uparrow +}
\nonumber\\
&&
\qquad \quad
+ e^{ip_{\uparrow -}z}(\hat\ve_{\uparrow -}\cdot\vec E_0)
\hat\ve_{\uparrow -}] .
\eea
Here,
$p_{\uparrow \pm} = [1-(\rh_\uparrow\pm\si_\uparrow)]p^0$,
where $\rh_\uparrow$ and $\si_\uparrow$ 
denote the values of $\rh$ and $\si$
for light with phase velocity in the $z$ direction.
The unit vectors $\hat\ve_ {\uparrow \pm}$
are the associated birefringent basis.

We suppose that the wave reflected at $z=l$
has phase velocity in the $-z$ direction.
Decomposing this wave in the same fashion gives
\bea
\vec E_1(x)&=&e^{-ip^0t}\bigl(
e^{-ip_{\downarrow +}z}(\hat\ve_{\downarrow +}\cdot\vec E_1)
\hat\ve_{\downarrow +}
\nonumber\\ 
&&
\qquad \quad
+ e^{-ip_{\downarrow -}z}(\hat\ve_{\downarrow -}\cdot\vec E_1)
\hat\ve_{\downarrow -}\bigr) ,
\eea
where the subscript $\downarrow$ denotes quantities 
for phase velocity in the $-z$ direction.
Similar expressions can be written for the electric field 
$\vec E_n(x)$ after $n$ reflections.
For the $n$th reflection with $n$ odd, 
the incident and reflected waves are related by 
$\vec E_{n+1}(x)|_{z=l} = e^{i\de_{l,n}}\vec E_n(x)|_{z=l}$.
A similar relation involving $\de_{0,n}$
holds for even $n$ at $z=0$.
The complex factors 
$e^{i\de_{0,n}}$, $e^{i\de_{l,n}}$
account for any phase change or loss 
due to transmission or absorption.
They may depend on the interaction of the wave
with the surfaces and 
could also depend on the incident polarization 
and various coefficients for Lorentz violation.
For simplicity, 
we suppose here that they are constant,
and denote them by $\de_0$ and $\de_l$. 

Superposing the contributions inside the cavity 
yields the total electric field as
\bea
\vec E(x) &=& e^{-ip^0t} [
( e^{ip_{\uparrow +}z}\hat\ve_{\uparrow +}\hat\ve_{\uparrow +}^\dag
+ e^{ip_{\uparrow -}z}\hat\ve_{\uparrow -}\hat\ve_{\uparrow -}^\dag )
\nonumber \\ 
&& 
\hskip -30pt
+ e^{-i\de_0}
( e^{-ip_{\downarrow +}z}\hat\ve_{\downarrow +}\hat\ve_{\downarrow +}^\dag
+ e^{-ip_{\downarrow -}z}\hat\ve_{\downarrow -}\hat\ve_{\downarrow -}^\dag)
m] 
\cdot M \cdot \vec E_0 .
\nonumber\\
\label{Ecavity}
\eea
At leading order,
the matrix $m$ is given by
\bea
m&=&e^{i(2p^0l+\de_0+\de_l)} \nonumber \\
&\times&\left[ 1-i2p^0l\left(
\begin{array}{ccc}
\overline{\rh+\si\cos\xi} & \overline{\si\sin\xi} & 0 \\
\overline{\si\sin\xi} & \overline{\rh-\si\cos\xi} & 0 \\
0&0&0
\end{array} \right)
\right] \ ,
\eea
where the bar signifies the average value
over the two different propagation directions.
The matrix $M$ is the geometric
series $M=\sum_{n=0}^\infty m^n = (1-m)^{-1}$.

The resonant frequency is often viewed as the frequency 
at which a standing wave is produced in the cavity.
However,
this notion may fail for nonzero $\kf$ 
because the wavelength of light traveling in one direction 
can differ from that for light traveling in the opposite direction.
A more appropriate definition
that applies also in the conventional case
is to take the resonant frequency as the frequency
maximizing the magnitude of the electric field
or the energy density.
The resonant frequency for a cavity is determined experimentally
by measuring the transmitted light,
so we adopt the energy density
of the transmitted light as the relevant quantity.
Using instead the magnitude 
of the transmitted electric or magnetic field
yields the same result at leading order.

We assume that the electric field $\vec E_T$
of the transmitted light is 
proportional to the component of the total
electric field in the cavity propagating in the $z$ direction:
\beq
\vec E_T \propto e^{-ip^0t}( 
e^{ip_{\uparrow +}z}\hat\ve_{\uparrow +}\hat\ve_{\uparrow +}^\dag
+ e^{ip_{\uparrow -}z}\hat\ve_{\uparrow -}\hat\ve_{\uparrow -}^\dag)
\cdot M \cdot \vec E_0 .
\eeq
The time-averaged energy density is
\bea
\expect u &=&\frac14 \Re(\vec E^* \cdot \vec D + \vec B^* \cdot \vec H)
\nonumber\\
&=&
\frac14 [\vec E^* \cdot (1+\ka_{DE}) \cdot \vec E
+ \vec B^* \cdot (1+\ka_{HB}) \cdot \vec B ].
\eea
With this equation and the Faraday law $ip^0 \vec B = \curl E$,
the energy density of the transmitted wave can be calculated.
Maximizing with respect to $p^0$ and solving for $p^0$
yields the perturbed resonant frequency.
We find 
\beq
\fr{\de\nu}{\nu} =
\fr{\vec E_0^*}{|\vec E_0|} \cdot \left(\begin{array}{ccc}
\overline{\rh+\si\cos\xi} & \overline{\si\sin\xi} & 0 \\
\overline{\si\sin\xi} & \overline{\rh-\si\cos\xi} & 0 \\
0&0&0
\end{array}\right) \cdot\fr{\vec E_0}{|\vec E_0|} \ .
\label{dnu2}
\eeq
The barred quantities can be determined from
\bea
\rh&=&-\half(\tilde\ka_{e-})^{11}-\half(\tilde\ka_{e-})^{22}
-(\tilde\ka_{o+})^{12}
-\tilde\ka_{\rm tr},
\nonumber\\
\si\sin\xi &=&\half(\tilde\ka_{o-})^{11}-\half(\tilde\ka_{o-})^{22}
-(\tilde\ka_{e+})^{12},
\nonumber\\
\si\cos\xi &=&-\half(\tilde\ka_{e+})^{11}+\half(\tilde\ka_{e+})^{22}
-(\tilde\ka_{o-})^{12},
\label{xi2}
\eea
which holds for a wave traveling in the $+z$ direction.
For a wave traveling in the $-z$ direction,
one must instead use Eq.\ \rf{xi2} with sign changes for the
parity-odd coefficients:
$\ka_{DB}\rightarrow-\ka_{DB}$, $\ka_{HE}\rightarrow-\ka_{HE}$.
The barred quantities in Eq.\ \rf{dnu} then are merely those 
in \rf{xi2} with $\ka_{DB}=\ka_{HE}=0$.
The net result is Eq.\ \rf{dnuopt}, as desired.

\section{Laboratory-frame quantities}
\label{signalcoeff}

In the scenario of Sec.\ \ref{optcavth}
and in terms of the matrices $\tilde\ka$ 
introduced in Eq.\ \rf{kappas2},
the quantities $A_{0,1,2,3,4}$ defined in Eq.\ \rf{dnu11}
can be written as
\bea
A_0&=&\frac18(1-3\cos^2\ch)
[3\ep_+(\tilde\ka_{e+})^{ZZ}
+\ep_-(\tilde\ka_{e-})^{ZZ}]
\nonumber \\
&&
-\frac18(9\ep_+-\ep_-)\tilde\ka_{\rm tr}
\nonumber \\
&&
+\frac14\be_\oplus
\bigl[2\bigl(\ep_--3\ep_+\cos^2\ch\bigr)
\sin\et\cos\Om_\oplus T
(\tilde\ka_{o+})^{XY}
\nonumber \\
&&+\bigl(3\ep_+-2\ep_--3\ep_+\cos^2\ch\bigr)
\nonumber \\
&&\times\bigl(\cos\et\cos\Om_\oplus T
(\tilde\ka_{o+})^{XZ}
+\sin\Om_\oplus T
(\tilde\ka_{o+})^{YZ}\bigr)
\nonumber \\
&&-3\ep_+(1-3\cos^2\ch)
\nonumber \\
&&\times\bigl(\cos\et\cos\Om_\oplus T
(\tilde\ka_{o-})^{XZ}
+\sin\Om_\oplus T
(\tilde\ka_{o-})^{YZ}\bigr)\bigr]
\nonumber \\
&&+\frac94\ep_+\be_L\sin\ch\cos\ch(\tilde\ka_{o-})^{ZZ},
\nonumber \\
A_1&=&-\frac12\sin\ch\cos\ch
[3\ep_+(\tilde\ka_{e+})^{YZ} +\ep_-(\tilde\ka_{e-})^{YZ}]
\nonumber \\
&&+\frac32\ep_+\be_\oplus\sin\ch\cos\ch
[\sin\Om_\oplus T
((\tilde\ka_{o-})^{YY}-(\tilde\ka_{o-})^{ZZ})
\nonumber \\
&&-\cos\et\cos\Om_\oplus T
((\tilde\ka_{o+})^{XY}-(\tilde\ka_{o-})^{XY})
\nonumber \\
&&+\sin\et\cos\Om_\oplus T
((\tilde\ka_{o+})^{XZ}-(\tilde\ka_{o-})^{XZ})]
\nonumber \\
&&+\frac12\be_L[\ep_-(\tilde\ka_{o+})^{YZ}
\nonumber \\
&&+3\ep_+(\sin^2\ch-\cos^2\ch)(\tilde\ka_{o-})^{YZ}],
\nonumber \\
A_2&=&-\frac12\sin\ch\cos\ch
[3\ep_+(\tilde\ka_{e+})^{XZ} +\ep_-(\tilde\ka_{e-})^{XZ}]
\nonumber \\
&&+\frac32\ep_+\be_\oplus\sin\ch\cos\ch
[\sin\Om_\oplus T
((\tilde\ka_{o+})^{XY}+(\tilde\ka_{o-})^{XY})
\nonumber \\
&&+\cos\et\cos\Om_\oplus T
((\tilde\ka_{o-})^{XX}-(\tilde\ka_{o-})^{ZZ})
\nonumber \\
&&-\sin\et\cos\Om_\oplus T
((\tilde\ka_{o+})^{YZ}-(\tilde\ka_{o-})^{YZ})]
\nonumber \\
&&+\frac12\be_L[\ep_-(\tilde\ka_{o+})^{XZ}
\nonumber \\
&&+3\ep_+(\sin^2\ch-\cos^2\ch)(\tilde\ka_{o-})^{XZ}],
\nonumber \\
A_3&=&-\frac14\sin^2\ch
[3\ep_+(\tilde\ka_{e+})^{XY} +\ep_-(\tilde\ka_{e-})^{XY}]
\nonumber \\
&&-\frac34\ep_+\be_\oplus\sin^2\ch
[\sin\Om_\oplus T
((\tilde\ka_{o+})^{XZ}+(\tilde\ka_{o-})^{XZ})
\nonumber \\
&&+\cos\et\cos\Om_\oplus T
((\tilde\ka_{o+})^{YZ}+(\tilde\ka_{o-})^{YZ})
\nonumber \\
&&+\sin\et\cos\Om_\oplus T
((\tilde\ka_{o-})^{XX}-(\tilde\ka_{o-})^{YY})]
\nonumber \\
&&-\frac32\ep_+\be_L\sin\ch\cos\ch
(\tilde\ka_{o-})^{XY},
\nonumber \\
A_4&=&-\frac18\sin^2\ch
\bigl[3\ep_+
\bigl((\tilde\ka_{e+})^{XX}
-(\tilde\ka_{e+})^{YY}\bigr)
\nonumber \\
&& +\ep_-
\bigl((\tilde\ka_{e-})^{XX}
-(\tilde\ka_{e-})^{YY}\bigr)\bigr]
\nonumber \\
&&+\frac34\ep_+\be_\oplus\sin^2\ch
[\sin\Om_\oplus T
((\tilde\ka_{o+})^{YZ}+(\tilde\ka_{o-})^{YZ})
\nonumber \\
&&-\cos\et\cos\Om_\oplus T
((\tilde\ka_{o+})^{XZ}+(\tilde\ka_{o-})^{XZ})
\nonumber \\
&&+2\sin\et\cos\Om_\oplus T
(\tilde\ka_{o-})^{XY}]
\nonumber \\
&&-\frac34\ep_+\be_L\sin\ch\cos\ch
[(\tilde\ka_{o-})^{XX}-(\tilde\ka_{o-})^{YY}].
\label{A}
\eea
In this equation,
we have introduced the quantities
\beq
\ep_+=\fr{2+\ep}{3\ep} , 
\quad
\ep_-=\fr{2-\ep}\ep ,
\eeq
which both reduce to 1 in the vacuum limit $\ep \to 1$.

The remaining coefficients are independent of $\ep$.
The coefficients $B_{0,1,2,3,4}$ are given by 
\bea
B_0&=&-\half\be_L\sin\ch(\tilde\ka_{o+})^{XY}
,
\nonumber \\
B_1&=&-\half\sin\ch
[(\tilde\ka_{e+})^{XZ} -(\tilde\ka_{e-})^{XZ}]
\nonumber \\
&&+\half\be_\oplus\sin\ch
[\sin\Om_\oplus T
((\tilde\ka_{o+})^{XY}+(\tilde\ka_{o-})^{XY})
\nonumber \\
&&+\cos\et\cos\Om_\oplus T
((\tilde\ka_{o-})^{XX}-(\tilde\ka_{o-})^{ZZ})
\nonumber \\
&&-\sin\et\cos\Om_\oplus T
((\tilde\ka_{o+})^{YZ}-(\tilde\ka_{o-})^{YZ})]
\nonumber \\
&&-\half\be_L\cos\ch
((\tilde\ka_{o+})^{XZ}+(\tilde\ka_{o-})^{XZ})
,
\nonumber \\
B_2&=&\half\sin\ch
[(\tilde\ka_{e+})^{YZ} -(\tilde\ka_{e-})^{YZ}]
\nonumber \\
&&-\half\be_\oplus\sin\ch
[\sin\Om_\oplus T
((\tilde\ka_{o-})^{YY}-(\tilde\ka_{o-})^{ZZ})
\nonumber \\
&&-\cos\et\cos\Om_\oplus T
((\tilde\ka_{o+})^{XY}-(\tilde\ka_{o-})^{XY})
\nonumber \\
&&+\sin\et\cos\Om_\oplus T
((\tilde\ka_{o+})^{XZ}-(\tilde\ka_{o-})^{XZ})]
\nonumber \\
&&+\half\be_L\cos\ch
((\tilde\ka_{o+})^{YZ}+(\tilde\ka_{o-})^{YZ})
,
\nonumber \\
B_3&=&\frac14\cos\ch
[(\tilde\ka_{e+})^{XX} -(\tilde\ka_{e-})^{XX}
\nonumber \\
&&-(\tilde\ka_{e+})^{YY} +(\tilde\ka_{e-})^{YY}]
\nonumber \\
&&-\half\be_\oplus\cos\ch
[\sin\Om_\oplus T
((\tilde\ka_{o+})^{YZ}+(\tilde\ka_{o-})^{YZ})
\nonumber \\
&&-\cos\et\cos\Om_\oplus T
((\tilde\ka_{o+})^{XZ}+(\tilde\ka_{o-})^{XZ})
\nonumber \\
&&+2\sin\et\cos\Om_\oplus T
(\tilde\ka_{o-})^{XY}]
\nonumber \\
&&-\frac14\be_L\sin\ch
[(\tilde\ka_{o-})^{XX}-(\tilde\ka_{o-})^{YY}]
,
\nonumber \\
B_4&=&-\half\cos\ch
[(\tilde\ka_{e+})^{XY} -(\tilde\ka_{e-})^{XY}]
\nonumber \\
&&-\half\be_\oplus\cos\ch
[\sin\Om_\oplus T
((\tilde\ka_{o+})^{XZ}+(\tilde\ka_{o-})^{XZ})
\nonumber \\
&&+\cos\et\cos\Om_\oplus T
((\tilde\ka_{o+})^{YZ}+(\tilde\ka_{o-})^{YZ})
\nonumber \\
&&+\sin\et\cos\Om_\oplus T
((\tilde\ka_{o-})^{XX}-(\tilde\ka_{o-})^{YY})]
\nonumber \\
&&+\half\be_L\sin\ch
(\tilde\ka_{o-})^{XY} .
\label{B}
\eea
The coefficients $C_{0,1,2,3,4}$ are 
\bea
C_0&=&-\frac38\sin^2\ch
[(\tilde\ka_{e+})^{ZZ} -(\tilde\ka_{e-})^{ZZ}]
\nonumber \\
&&-\frac14\be_\oplus\sin^2\ch
[\sin\Om_\oplus T
((\tilde\ka_{o+})^{YZ}-3(\tilde\ka_{o-})^{YZ})
\nonumber \\
&&+\cos\et\cos\Om_\oplus T
((\tilde\ka_{o+})^{XZ}-3(\tilde\ka_{o-})^{XZ})
\nonumber \\
&&+2\sin\et\cos\Om_\oplus T
(\tilde\ka_{o+})^{XY}]
\nonumber \\
&&-\frac34\be_L\sin\ch\cos\ch
(\tilde\ka_{o-})^{ZZ}
,
\nonumber \\
C_1&=&\half\sin\ch\cos\ch
[(\tilde\ka_{e+})^{YZ} -(\tilde\ka_{e-})^{YZ}]
\nonumber \\
&&-\half\be_\oplus\sin\ch\cos\ch
[\sin\Om_\oplus T
((\tilde\ka_{o-})^{YY}-(\tilde\ka_{o-})^{ZZ})
\nonumber \\
&&-\cos\et\cos\Om_\oplus T
((\tilde\ka_{o+})^{XY}-(\tilde\ka_{o-})^{XY})
\nonumber \\
&&+\sin\et\cos\Om_\oplus T
((\tilde\ka_{o+})^{XZ}-(\tilde\ka_{o-})^{XZ})]
\nonumber \\
&&+\half\be_L
[(\tilde\ka_{o+})^{YZ}
-(\sin^2\ch-\cos^2\ch)(\tilde\ka_{o-})^{YZ}]
,
\nonumber \\
C_2&=&\half\sin\ch\cos\ch
[(\tilde\ka_{e+})^{XZ} -(\tilde\ka_{e-})^{XZ}]
\nonumber \\
&&-\half\be_\oplus\sin\ch\cos\ch
[\sin\Om_\oplus T
((\tilde\ka_{o+})^{XY}+(\tilde\ka_{o-})^{XY})
\nonumber \\
&&+\cos\et\cos\Om_\oplus T
((\tilde\ka_{o-})^{XX}-(\tilde\ka_{o-})^{ZZ})
\nonumber \\
&&-\sin\et\cos\Om_\oplus T
((\tilde\ka_{o+})^{YZ}-(\tilde\ka_{o-})^{YZ})]
\nonumber \\
&&+\half\be_L
[(\tilde\ka_{o+})^{XZ}
-(\sin^2\ch-\cos^2\ch)(\tilde\ka_{o-})^{XZ}]
,
\nonumber \\
C_3&=&-\frac14(1+\cos^2\ch)
[(\tilde\ka_{e+})^{XY} -(\tilde\ka_{e-})^{XY}]
\nonumber \\
&&-\frac14\be_\oplus(1+\cos^2\ch)
[\sin\Om_\oplus T
((\tilde\ka_{o+})^{XZ}+(\tilde\ka_{o-})^{XZ})
\nonumber \\
&&+\cos\et\cos\Om_\oplus T
((\tilde\ka_{o+})^{YZ}+(\tilde\ka_{o-})^{YZ})
\nonumber \\
&&+\sin\et\cos\Om_\oplus T
((\tilde\ka_{o-})^{XX}-(\tilde\ka_{o-})^{YY})]
\nonumber \\
&&+\half\be_L\sin\ch\cos\ch (\tilde\ka_{o-})^{XY}
,
\nonumber \\
C_4&=&-\frac18(1+\cos^2\ch)
[(\tilde\ka_{e+})^{XX} -(\tilde\ka_{e-})^{XX}
\nonumber \\
&&-(\tilde\ka_{e+})^{YY} +(\tilde\ka_{e-})^{YY}]
\nonumber \\
&&+\frac14\be_\oplus(1+\cos^2\ch)
[\sin\Om_\oplus T
((\tilde\ka_{o+})^{YZ}+(\tilde\ka_{o-})^{YZ})
\nonumber \\
&&-\cos\et\cos\Om_\oplus T
((\tilde\ka_{o+})^{XZ}+(\tilde\ka_{o-})^{XZ})
\nonumber \\
&&+2\sin\et\cos\Om_\oplus T
(\tilde\ka_{o-})^{XY}]
\nonumber \\
&&+\frac14\be_L \sin\ch\cos\ch
[(\tilde\ka_{o-})^{XX}-(\tilde\ka_{o-})^{YY}] .
\label{C}
\eea

\section{Satellite-frame quantities}
\label{spacecoeff}

The quantities ${\cal A}_{s,c}$, ${\cal B}_{s,c}$
appearing in Eq.\ \rf{dnu7} of Sec.\ \ref{spaceexpt}
can be expressed in terms of 
the matrices $\tilde\ka$ introduced in Eq.\ \rf{kappas2}
through the convenient combinations \rf{convenient}.
In terms of the various orientation angles
specified in Appendix \ref{frames},
for the quantity ${\cal A}_s$ we find
\bea
{\cal A}_s&=& \frac14\cos2\ze
[\sin\al(\tilde\ka_{e'})^{XZ}
-\cos\al(\tilde\ka_{e'})^{YZ}]
\nonumber\\
&&
+\frac18\sin2\ze
[(1+\sin^2\al)(\tilde\ka_{e'})^{XX}
\nonumber\\
&&
+(1+\cos^2\al)(\tilde\ka_{e'})^{YY}
-\sin2\al(\tilde\ka_{e'})^{XY}]
\nonumber\\
&&
+\frac14\be_s
\bigl[\sin\al\bigl((\tilde\ka_{o'})^{XZ}
-(\tilde\ka_{o'})^{ZX}\bigr)
\nonumber\\
&&
-\cos\al\bigl((\tilde\ka_{o'})^{YZ}
-(\tilde\ka_{o'})^{ZY}\bigr)\bigr]
\nonumber\\
&&
+\frac14\be_\oplus\bigl[
\cos\al\cos2\ze\cos\Om_\oplus T
\nonumber\\
 &&
\times\bigl(\cos\et
(\tilde\ka_{o'})^{YX}
-\sin\et(\tilde\ka_{o'})^{ZX}\bigr)
\nonumber\\
&&
+\cos\al\cos2\ze\sin\Om_\oplus T
\bigl( (\tilde\ka_{o'})^{YY} -(\tilde\ka_{o'})^{ZZ} \bigr) 
\nonumber\\
&&
-\sin\al\cos2\ze\cos\Om_\oplus T
\nonumber\\
 &&
\times\bigl(\cos\et
\bigl((\tilde\ka_{o'})^{XX}
-(\tilde\ka_{o'})^{ZZ}\bigr)
+\sin\et(\tilde\ka_{o'})^{ZY}\bigr)
\nonumber\\
&&
-\sin\al\cos2\ze\sin\Om_\oplus T
(\tilde\ka_{o'})^{XY}
\nonumber\\
&&
+\cos^2\al\sin2\ze\cos\Om_\oplus T
\nonumber\\
 &&
\times\bigl(\cos\et
(\tilde\ka_{o'})^{ZX}
+\sin\et(\tilde\ka_{o'})^{YX}\bigr)
\nonumber\\
&&
+\cos^2\al\sin2\ze\sin\Om_\oplus T
\bigl((\tilde\ka_{o'})^{YZ}
+(\tilde\ka_{o'})^{ZY}\bigr)
\nonumber\\
&&
+\sin^2\al\sin2\ze\cos\Om_\oplus T
\nonumber\\
 &&
\times\bigl(\cos\et
\bigl((\tilde\ka_{o'})^{XZ}
+(\tilde\ka_{o'})^{ZX}\bigr)
-\sin\et(\tilde\ka_{o'})^{XY}\bigr)
\nonumber\\
&&
+\sin^2\al\sin2\ze\sin\Om_\oplus T
(\tilde\ka_{o'})^{ZY}
\nonumber\\
&&
-\half\sin2\al\sin2\ze\cos\Om_\oplus T
\nonumber\\
&&
\times\bigl(\cos\et
(\tilde\ka_{o'})^{YZ}
+\sin\et\bigl((\tilde\ka_{o'})^{XX}
-(\tilde\ka_{o'})^{YY}\bigr)\bigr)
\nonumber\\
&&
-\half\sin2\al\sin2\ze\sin\Om_\oplus T
(\tilde\ka_{o'})^{XZ}
\bigr].
\eea
The quantity ${\cal A}_c$ is
\bea
{\cal A}_c&=&
-\frac14\cos\ze\sin\al(\tilde\ka_{e'})^{YZ}
\nonumber\\
&&
-\frac14\cos\ze\cos\al(\tilde\ka_{e'})^{XZ}
\nonumber\\
&&
-\frac18\sin\ze\sin2\al
\bigl((\tilde\ka_{e'})^{XX}
-(\tilde\ka_{e'})^{YY}\bigr)
\nonumber\\
&&
+\frac14\sin\ze\cos2\al
(\tilde\ka_{e'})^{XY}
\nonumber\\
&&
+\frac14\be_s\bigl[
\sin\ze
\bigl((\tilde\ka_{o'})^{XY}
-(\tilde\ka_{o'})^{YX}\bigr)
\nonumber\\
&&
+\cos\ze\cos\al
\bigl((\tilde\ka_{o'})^{ZX}
-(\tilde\ka_{o'})^{XZ}\bigr)
\nonumber\\
&&
+\cos\ze\sin\al
\bigl((\tilde\ka_{o'})^{ZY}
-(\tilde\ka_{o'})^{YZ}\bigr)\bigr]
\nonumber\\
&&
+\frac14\be_\oplus\bigl[
\cos\al\cos\ze\cos\Om_\oplus T
\nonumber\\
 &&
\times\bigl(\cos\et\bigl(
(\tilde\ka_{o'})^{XX}
-(\tilde\ka_{o'})^{ZZ}\bigr)
+\sin\et(\tilde\ka_{o'})^{ZY}\bigr)
\nonumber\\
&&
+\cos\al\cos\ze\sin\Om_\oplus T
(\tilde\ka_{o'})^{XY}
\nonumber\\
&&
+\sin\al\cos\ze\cos\Om_\oplus T
\nonumber\\
 &&
\times\bigl(\cos\et
(\tilde\ka_{o'})^{YX}
-\sin\et(\tilde\ka_{o'})^{ZX}\bigr)
\nonumber\\
&&
+\sin\al\cos\ze\sin\Om_\oplus T
\bigl((\tilde\ka_{o'})^{YY}
-(\tilde\ka_{o'})^{ZZ}\bigr)
\nonumber\\
&&
+\cos2\al\sin\ze\cos\Om_\oplus T
\nonumber\\
 &&
\times\bigl(\cos\et
(\tilde\ka_{o'})^{YZ}
+\sin\et\bigl((\tilde\ka_{o'})^{XX}
-(\tilde\ka_{o'})^{YY}\bigr)\bigr)
\nonumber\\
&&
+\cos2\al\sin\ze\sin\Om_\oplus T
(\tilde\ka_{o'})^{XZ}
\nonumber\\
&&
-\sin2\al\sin\ze\cos\Om_\oplus T
\nonumber\\
 &&
\times\bigl(\cos\et
(\tilde\ka_{o'})^{XZ}
-\sin\et\bigl((\tilde\ka_{o'})^{XY}
+(\tilde\ka_{o'})^{YX}\bigr)\bigr)
\nonumber\\
&&
+\sin2\al\sin\ze\sin\Om_\oplus T
(\tilde\ka_{o'})^{YZ}
\bigr].
\eea
The quantity ${\cal B}_s$ is
\bea
{\cal B}_s&=&
\frac38\sin\ze\sin\al(\tilde\ka_{e'})^{YZ}
+\frac38\sin\ze\cos\al(\tilde\ka_{e'})^{XZ}
\nonumber\\
&&
-\frac3{16}\cos\ze\sin2\al\bigl(
(\tilde\ka_{e'})^{XX}
-(\tilde\ka_{e'})^{YY}\bigr)
\nonumber\\
&&
+\frac38\cos\ze\cos2\al
(\tilde\ka_{e'})^{XY}
\nonumber\\
&&
-\frac18\be_s\bigl[
\sin\ze\cos\al
\bigl((\tilde\ka_{o'})^{XZ}
+(\tilde\ka_{o'})^{ZX}\bigr)
\nonumber\\
&&
+\sin\ze\sin\al
\bigl((\tilde\ka_{o'})^{YZ}
+(\tilde\ka_{o'})^{ZY}\bigr)
\nonumber\\
&&
-\cos\ze\sin2\al
\bigl((\tilde\ka_{o'})^{XX}
-(\tilde\ka_{o'})^{YY}\bigr)
\nonumber\\
&&
+\cos\ze\cos2\al
\bigl((\tilde\ka_{o'})^{XY}
+(\tilde\ka_{o'})^{YX}\bigr)\bigr]
\nonumber\\
&&
+\frac38\be_\oplus\bigl[
-\cos\al\sin\ze\cos\Om_\oplus T
\nonumber\\
 &&
\times\bigl(\cos\et\bigl(
(\tilde\ka_{o'})^{XX}
-(\tilde\ka_{o'})^{ZZ}\bigr)
+\sin\et(\tilde\ka_{o'})^{ZY}\bigr)
\nonumber\\
&&
-\cos\al\sin\ze\sin\Om_\oplus T
(\tilde\ka_{o'})^{XY}
\nonumber\\
&&
-\sin\al\sin\ze\cos\Om_\oplus T
\nonumber\\
 &&
\times\bigl(\cos\et
(\tilde\ka_{o'})^{YX}
-\sin\et(\tilde\ka_{o'})^{ZX}\bigr)
\nonumber\\
&&
-\sin\al\sin\ze\sin\Om_\oplus T
\bigl((\tilde\ka_{o'})^{YY}
-(\tilde\ka_{o'})^{ZZ}\bigr)
\nonumber\\
&&
+\cos2\al\cos\ze\cos\Om_\oplus T
\nonumber\\
 &&
\times\bigl(\cos\et
(\tilde\ka_{o'})^{YZ}
+\sin\et\bigl((\tilde\ka_{o'})^{XX}
-(\tilde\ka_{o'})^{YY}\bigr)\bigr)
\nonumber\\
&&
+\cos2\al\cos\ze\sin\Om_\oplus T
(\tilde\ka_{o'})^{XZ}
\nonumber\\
&&
-\sin2\al\cos\ze\cos\Om_\oplus T
\nonumber\\
 &&
\times\bigl(\cos\et
(\tilde\ka_{o'})^{XZ}
-\sin\et\bigl((\tilde\ka_{o'})^{XY}
+(\tilde\ka_{o'})^{YX}\bigr)\bigr)
\nonumber\\
&&
+\sin2\al\cos\ze\sin\Om_\oplus T
(\tilde\ka_{o'})^{YZ}
\bigr].
\eea
The quantity ${\cal B}_c$ is
\bea
{\cal B}_c&=&\frac3{16}(\cos^2\al
-\sin^2\al\cos^2\ze+\sin^2\ze)
(\tilde\ka_{e'})^{XX}
\nonumber\\
&&
+\frac3{16}(\sin^2\al
-\cos^2\al\cos^2\ze+\sin^2\ze)
(\tilde\ka_{e'})^{YY}
\nonumber\\
&&
+\frac3{16}\sin2\al
(1+\cos^2\ze)
(\tilde\ka_{e'})^{XY}
\nonumber\\
&&
+\frac3{16}\sin2\ze
[\sin\al(\tilde\ka_{e'})^{XZ}
-\cos\al(\tilde\ka_{e'})^{YZ}]
\nonumber\\
&&
-\frac1{16}\be_s\bigl[
2(\cos^2\al
-\sin^2\al\cos^2\ze+\sin^2\ze)
(\tilde\ka_{o'})^{XX}
\nonumber\\
&&
+2(\sin^2\al
-\cos^2\al\cos^2\ze+\sin^2\ze)
(\tilde\ka_{o'})^{YY}
\nonumber\\
&&
+\sin2\al(2\cos^2\ze+\sin^2\ze)
\bigl((\tilde\ka_{o'})^{XY}
+(\tilde\ka_{o'})^{YX}\bigr)
\nonumber\\
&&
+\sin\al\sin2\ze
\bigl((\tilde\ka_{o'})^{XZ}
+(\tilde\ka_{o'})^{ZX}\bigr)
\nonumber\\
&&
-\cos\al\sin2\ze
\bigl((\tilde\ka_{o'})^{YZ}
+(\tilde\ka_{o'})^{ZY}\bigr)\bigr]
\nonumber\\
&&
+\frac3{16}\be_\oplus\bigl[
\sin2\al(1+\cos^2\ze)\cos\Om_\oplus T
\nonumber\\
&&
\times\bigl(
\cos\et(\tilde\ka_{o'})^{YZ}
+\sin\et\bigl(
(\tilde\ka_{o'})^{XX}
-(\tilde\ka_{o'})^{YY}\bigr)\bigr)
\nonumber\\
&&
+\sin2\al(1+\cos^2\ze)\sin\Om_\oplus T
(\tilde\ka_{o'})^{XZ}
\nonumber\\
&&
-\sin\al\sin2\ze\cos\Om_\oplus T
\nonumber\\
&&
\times\bigl(\cos\et
\bigl((\tilde\ka_{o'})^{XX}
-(\tilde\ka_{o'})^{ZZ}\bigr)
+\sin\et(\tilde\ka_{o'})^{ZX}\bigr)
\nonumber\\
&&
-\sin\al\sin2\ze\sin\Om_\oplus T
(\tilde\ka_{o'})^{XY}
\nonumber\\
&&
+\cos\al\sin2\ze\cos\Om_\oplus T
\nonumber\\
&&
\times\bigl(\cos\et
(\tilde\ka_{o'})^{YX}
-\sin\et(\tilde\ka_{o'})^{ZX}\bigr)
\nonumber\\
&&
+\cos\al\sin2\ze\sin\Om_\oplus T
\bigl((\tilde\ka_{o'})^{YY}
-(\tilde\ka_{o'})^{ZZ}\bigr)
\nonumber\\
&&
-2(\sin^2\al\cos^2\ze-\cos^2\al)
\cos\Om_\oplus T
\nonumber\\
&&
\times\bigl(\cos\et
(\tilde\ka_{o'})^{XZ}
-\sin\et(\tilde\ka_{o'})^{XY}\bigr)
\nonumber\\
&&
-2(\cos^2\al\cos^2\ze-\sin^2\al)
\cos\Om_\oplus T\sin\et
(\tilde\ka_{o'})^{YX}
\nonumber\\
&&
-2(\cos^2\al\cos^2\ze-\sin^2\al)
\sin\Om_\oplus T
(\tilde\ka_{o'})^{YZ}
\nonumber\\
&&
+2\sin^2\ze\bigl(\cos\Om_\oplus T
\cos\et(\tilde\ka_{o'})^{ZX}
\nonumber\\
&&
\qquad
+\sin\Om_\oplus T
(\tilde\ka_{o'})^{ZY}\bigr)\bigr].
\eea
Finally, the quantity ${\cal C}$ is
\bea
{\cal C}&=&
\frac1{16}(3\sin^2\al\sin^2\ze-1)
(\tilde\ka_{e'})^{XX}
\nonumber\\
&&
+\frac1{16}(3\cos^2\al\sin^2\ze-1)
(\tilde\ka_{e'})^{YY}
\nonumber\\
&&
+\frac1{16}(3\cos^2\ze-1)
(\tilde\ka_{e'})^{ZZ}
+\frac3{16}\sin2\al\sin^2\ze
(\tilde\ka_{e'})^{XY}
\nonumber\\
&&
-\frac3{16}\sin2\ze
[\sin\al(\tilde\ka_{e'})^{XZ}
-\cos\al(\tilde\ka_{e'})^{YZ}]
\nonumber\\
&&
+\frac18\be_s\bigl[
(3\sin^2\al\sin^2\ze-1)
(\tilde\ka_{o'})^{XX}
\nonumber\\
&&
+(3\cos^2\al\sin^2\ze-1)
(\tilde\ka_{o'})^{YY}
\nonumber\\
&&
+(3\cos^2\ze-1)
(\tilde\ka_{o'})^{ZZ}
\nonumber\\
&&
-\frac32\sin2\al\sin^2\ze
\bigl((\tilde\ka_{o'})^{XY}
+(\tilde\ka_{o'})^{YX}\bigr)
\nonumber\\
&&
+\frac32\sin2\ze
\bigl(\sin\al\bigl(
(\tilde\ka_{o'})^{XZ}
+(\tilde\ka_{o'})^{ZX}\bigr)
\nonumber\\
&&
-\cos\al\bigl(
(\tilde\ka_{o'})^{YZ}
+(\tilde\ka_{o'})^{ZY}\bigr)\bigr)\bigr]
\nonumber\\
&&
+\frac18\be_\oplus\bigl[
(3\sin^2\al\sin^2\ze-1)
\cos\Om_\oplus T
\nonumber\\
&&
\times\bigl(\cos\et
(\tilde\ka_{o'})^{XZ}
-\sin\et
(\tilde\ka_{o'})^{XY}\bigl)
\nonumber\\
&&
+(3\cos^2\al\sin^2\ze-1)
\bigl(\cos\Om_\oplus T \sin\et
(\tilde\ka_{o'})^{YX}
\nonumber\\
&&
+\sin\Om_\oplus T
(\tilde\ka_{o'})^{YZ}\bigr)
\nonumber\\
&&
-(3\cos^2\ze-1)
\bigl(\cos\Om_\oplus T \cos\et
(\tilde\ka_{o'})^{ZX}
\nonumber\\
&&
+\sin\Om_\oplus T
(\tilde\ka_{o'})^{ZY}\bigr)
\nonumber\\
&&
-\frac32\sin2\al\sin^2\ze
\cos\Om_\oplus T \bigl(\cos\et
(\tilde\ka_{o'})^{YZ}
\nonumber\\
&&
+\sin\et\bigl(
(\tilde\ka_{o'})^{XX}
-(\tilde\ka_{o'})^{YY}\bigr)\bigr)
\nonumber\\
&&
-\frac32\sin2\al\sin^2\ze
\sin\Om_\oplus T
(\tilde\ka_{o'})^{XZ}
\nonumber\\
&&
-\frac32\sin\al\sin2\ze
\cos\Om_\oplus T \bigl(\cos\et
\bigl((\tilde\ka_{o'})^{XX}
-(\tilde\ka_{o'})^{ZZ}\bigr)
\nonumber\\
&&
+\sin\et(\tilde\ka_{o'})^{ZY}\bigr)
\nonumber\\
&&
-\frac32\sin\al\sin2\ze
\sin\Om_\oplus T
(\tilde\ka_{o'})^{XY}
\nonumber\\
&&
+\frac32\cos\al\sin2\ze
\cos\Om_\oplus T \bigl(\cos\et
(\tilde\ka_{o'})^{YX}
\nonumber\\
&&
-\sin\et(\tilde\ka_{o'})^{ZX}\bigr)
\nonumber\\
&&
+\frac32\cos\al\sin2\ze
\sin\Om_\oplus T\bigl(
(\tilde\ka_{o'})^{YY}
-(\tilde\ka_{o'})^{ZZ}\bigr)
\bigr].
\eea

\end{multicols}
\end{document}